# ARTICLE

# Influence of pore-confined water on the thermal expansion of a zinc-based metal-organic framework

Nina Strasser,[a] Benedikt Schrode,[b] Ana Torvisco,[c] Sanjay John,[a] Birgit Kunert,[a] Brigitte Bitschnau,[d] Florian Patrick Lindner,[a] Christian Slugovc,[e] Egbert Zojer*[a] and Roland Resel*[a]

Understanding the reversible intercalation of guest molecules into metal-organic frameworks is crucial for advancing their design for practical applications. In this work, we explore the impact of $H_2O$ as a guest molecule on the thermal expansion of the zinc-based metal-organic framework GUT-2. Dehydration is achieved by thermal treatment of hydrated GUT-2. Rietveld refinement performed on temperature-dependent X-ray powder diffraction data confirms the reversible structural transformation. Additionally, it allows the determination of anisotropic thermal expansion coefficients for both phases. The hydrated form exhibits near-zero thermal expansion along the polymer chain direction, moderate expansion In the direction of predominantly hydrogen bonds, and the highest expansion in the direction with only Van der Waals bonding. Upon activation, the removal of $H_2O$ molecules triggers a doubling of the thermal expansion coefficient in the direction, where the hydrogen bonds have been removed. Regarding the dynamics of the process, thermal activation in air occurs within 6 hours at a temperature of 50°C and takes only 30 minutes when heating to 90°C. In contrast, full rehydration under standard lab conditions (30 % relative humidity) requires two days. During the activation/dehydration processes no change of the widths of the X-ray diffraction peaks is observed, which shows that the underlying crystal structures remains fully intact during the transition processes. Fitting the transformations by the Avrami equation reveals a quasi one-dimensional evolution of the dehydrated areas for the activation process and a more intricate, predominantly two-dimensional mechanism for the rehydration.

## 1. Introduction

According to the IUPAC recommendation, a metal-organic framework (MOF) is a "coordination network with organic ligands containing potential voids".[1] These voids (often also called pores) can trap small molecules, a process that can fundamentally change the functionality of the MOF. In catalysis, for instance, MOFs can encapsulate molecular catalysts, shielding them from reactive species and preventing deactivation during the chemical reaction.[2–4] Moreover, MOFs can encapsulate enzymes, preventing their denaturation,[5] or can hold drug molecules, enabling their controlled release for maximizing the therapeutic responses.[6] MOFs are also promising candidates for electronic sensors, in which they serve as adsorption systems for gas molecule detection.[7,8]

Interaction of MOFs with small molecules can, however, pose significant risks to their structural integrity. For example, several systems belonging to the isoreticular MOF family[9] (which is a series of MOFs that have a similar network topology) including MOF-5[10,11] disintegrate after minimal exposure to $H_2O$. The primary reason is that $H_2O$ molecules can easily hydrolyse the relatively weak metal-organic coordination bonds. This then leads to the structural collapse of the whole framework. However, there are also MOFs which show exceptional $H_2O$ resistance, a feature that aligns with the recent development of innovative MOFs that allow harvesting $H_2O$ from atmospheric moisture, even amidst desert conditions.[12,13]

Generally speaking, $H_2O$ adsorption in MOFs involves interactions that vary in strength depending on the binding sites and the specific framework structure.[14] In some cases, $H_2O$ is weakly bound through hydrogen bonding, allowing for reversible adsorption and desorption, such as in the aforementioned $H_2O$-harvesting applications.[15] Conversely, certain MOFs exhibit binding sites, where $H_2O$ molecules are so strongly coordinated (e.g., via metal centres or specific functional groups) that $H_2O$ desorption becomes rather challenging.[16] These differences in binding strengths influence the structural responses of the framework to both $H_2O$ inclusion or desorption, leading to diverse effects that can range from simple expansions or contractions of the framework to more complex phenomena, such as ligand rearrangements or phase transitions. These responses have direct crystallographic

[a.] *Institute of Solid State Physics, NAWI Graz, Graz University of Technology, Petersgasse 16, 8010 Graz, Austria. E-mails: egbert.zojer@tugraz.at and roland.resel@tugraz.at*
[b.] *Anton Paar GmbH, Anton-Paar-Straße 20, 8054 Graz, Austria*
[c.] *Institute of Inorganic Chemistry, NAWI Graz, Graz University of Technology, Stremayrgasse 9, 8010 Graz, Austria*
[d.] *Institute of Physical and Theoretical Chemistry, NAWI Graz, Graz University of Technology, Stremayrgasse 9, 8010 Graz, Austria*
[e.] *Institute of Chemistry and Technology of Materials, NAWI Graz, Graz University of Technology, Stremayrgasse 9, 8010 Graz, Austria*
† Electronic supplementary information (ESI) available.
See DOI: 10.1039/x0xx00000x

consequences, such as modifications of cell parameters, changes in symmetry, or the reorientation of linkers.[17]

Alternative processes modifying the details of MOF structures in response to temperature changes are thermal expansion processes. In MOFs they occur in (a combination of) three flavours: (i) The most common response to changes in temperature is a positive thermal expansion (PTE), where the cell dimensions expand upon heating.[18–20] (ii) A less frequent phenomenon is negative thermal expansion (NTE), where cell dimensions decrease with increasing temperature (at least in one of the crystallographic directions).[21–23] (iii) Finally, a material may exhibit zero thermal expansion (ZTE), where the cell dimensions remain essentially unchanged with temperature.[24] The condition for ZTE is that one thermal expansion coefficient is smaller than $1\times10^{-6}$ K$^{-1}$.[25] ZTE is, for example, highly desirable in the design of materials for high-precision devices, where maintaining a constant shape and size across multiple temperature ranges is crucial for maintaining the accuracy of the device.[26]

Interestingly, it has been shown that the aforementioned adsorption of guest molecules can be used to control thermal expansion processes.[27–29] For instance, a lanthanide-based MOF incorporating dimethylformamide (DMF) molecules within its pores exhibited tuneable NTE attributed to a reduction in pore sizes upon guest molecule removal.[30] Another study on $Zn_2(BDC)_2(DABCO)$ compared the thermal expansion of the MOF containing DMF molecules with that of the system containing benzene as guest molecules.[31] This revealed distinct host-guest interactions likely responsible for varied thermal responses. In yet another case, the MOF PCN-222 along one axis displayed a thermal expansion coefficient that changed its sign depending on the $H_2O$ content within its pores.[32] While these previous studies have explored the role of guest molecules in influencing the thermal expansion of MOFs, a comprehensive understanding of how specific bonding interactions and structural rearrangements contribute to anisotropic thermal expansion remains limited – a gap this study aims to address.

Studying the intimate interplay between thermal expansion and guest adsorption is in the focus of the current manuscript. The experiments are performed on GUT-2,[33] a Zn-based MOF recently developed at Graz University of Technology. It has been chosen for this study for a variety of reasons: (i) GUT-2 is stable in $H_2O$ and humid environments, (ii) it can be activated and rehydrated without framework degradation in a reversible process, (iii) it can be rather straightforwardly grown into comparably large single crystals, which allows the exact determination of its atomistic structure, and (iv) it has a well-defined binding site for $H_2O$ molecules, at which they can form linkages between strands of the GUT-2 coordination polymer via establishing hydrogen bonds.

## 2. Experimental details

### 2.1. Synthesis

GUT-2 was synthesized in an aqueous solution following the protocol reported in Ref.[33]. After synthesis, the obtained white powder was dried by blowing a stream of $N_2$ gas over it.

### 2.2. Single crystal X-ray diffraction

The measurements of GUT-2 were performed on a Rigaku XtaLAB Synergy, Dualflex, HyPix-Arc 100 diffractometer equipped with an Oxford Cryosystems cryostream. A single crystal was selected from the as-synthesized (hydrated) powder and carefully mounted on a glass rod on a metal pin and secured with glue (Loctite Super Attack) to ensure that the single crystal does not drop off during the measurement. The glue was pre-tested to confirm that it contained no crystalline components, which would interfere with the diffraction experiment. Data were collected at -173°C (100 K) for hydrated GUT-2. The measurements of activated GUT-2 (373 K) were also performed directly on the instrument at an elevated temperature of 100°C, utilizing the cryostream with activation achieved in situ starting from the hydrated single crystal.

For the measurement, Cu Kα radiation (λ = 1.54056 Å) was used. The diffraction pattern was indexed and the total number of runs and images was based on the data collection strategy calculated by the program CrysAlisPro[34] (with all details available in the provided CIFs). The unit cell was refined and data reduction, scaling, and absorption corrections were performed with this program. Using OLEX2,[35] the structure was solved with the SHELXT structure solution program[36] and refined with the SHELXL refinement package[37] using full matrix least squares minimization on F2. All non-hydrogen atoms were refined anisotropically. Hydrogen atom positions were calculated geometrically and refined using the riding model. Activated GUT-2 (373 K) was refined as a 2-component twin (BASF 0.39). It showed comparably weak and diffuse Bragg scattering, which is related to the fact that the structure was collected at 373 K, which contributes to enlarged displacement parameters, indicative of an increased thermal motion of the atoms in accordance with data collection at higher temperature (see Supporting Information for more details).

CIF files were edited, validated and formatted either with the programs enCIFer,[38] publCIF[39] or OLEX2[35]. CCDC 2406179-2406180 contain the supplementary crystallographic data for fully hydrated GUT-2 (compound (1)) and for activated GUT-2 (compound (2)), respectively. These data can be obtained free of charge from The Cambridge Crystallographic Data Centre via www.ccdc.cam.ac.uk/data_request/cif. Table S1 contains crystallographic data and details of measurements and refinements for compounds (1) and (2).

### 2.3. Powder X-Ray diffraction – thermal expansion

Powder X-ray diffraction (PXRD) experiments for determining thermal expansion coefficients were performed with the XRDynamic 500 diffractometer from Anton Paar equipped with a Primux 3000 sealed-tube X-ray source with Cu anode (λ =

1.5418 Å) and a Pixos 2000 detection unit featuring a solid-state pixel detector with Si sensors operated in 1D mode. Data were collected in Bragg-Brentano beam geometry with a primary flat multilayer X-ray mirror from 10° to 50° 2θ with a step size of 0.01°. Soller slits with an opening of 0.05 rad were used on the primary and secondary side.

The cooling of the finely-grinded, as-synthesized GUT-2 sample was achieved using the low-temperature attachment TTK600 from Anton Paar Ltd. Graz under vacuum conditions ($1\times10^{-3}$ mbar) using liquid $N_2$ as a cooling agent. Starting from 25°C the sample was cooled/heated in steps of 50°C with a cooling/heating rate of 2°C/min. Waiting times and automatic sample alignments were applied at each temperature step to ensure homogeneous sample temperature and sample alignment. During the refilling of liquid $N_2$ at 25°C, moisture infiltrated into the supply line. This moisture subsequently froze, leading to a reduction in the cooling efficiency during the second cooling run, limiting the lowest achievable temperature to -180°C.

In order to obtain the anisotropic thermal expansion coefficients from the temperature-dependent PXRD patterns, Rietveld refinements[40] were performed using the program X'Pert Highscore Plus[41] (Version: 3.0). The structural models used for the refinements were the crystal structure solutions obtained by single-crystal XRD experiments. The profile parameters that were included in the refinements are the zero shift, an overall scale factor, the three cell lengths as well as the line broadening parameters U, V and W of the Caglioti function.[42–44] Moreover, the background contributions were modelled using a polynomial function, while the peak-shape function is represented by a pseudo-Voigt function. The diffracted intensities were corrected for preferred orientation effects, which indicated that the (010) plane in the hydrated form is preferentially aligned parallel to the sample holder base.[45] For the activated from it is the orientation stays the same, but when applying the Niggli naming convention[46] for the crystallographic directions, this means that now the (001) plane would be aligned parallel (see discussion below). The full width at half maximum (FWHM) is then calculated as the peak width where the peak intensity falls to fifty percent of its maximum value.

**2.4. Powder X-ray diffraction – activation and hydration kinetics**

Temperature-dependent PXRD measurements for studying the kinetics of activation and rehydration were performed on a PANalytical Empyrean system in combination with a sealed copper tube and using a DHS900 heating attachment[47] from Anton Paar. For all powder diffraction experiments, GUT-2 powder finely grinded in a porcelain mortar was put on a silicon wafer. The primary X-ray beam was monochromatized and parallelized by an X-ray mirror. The diffracted beam was detected by a 1-dimensional detector mode using a PIXcel3D detector with 255 active channels. In this mode, the detector simultaneously records diffraction intensities along a single axis (2θ). An anti-scatter slit of 7.5 mm and a 0.02 rad Soller slit were used. Careful alignments of the sample height were performed at each temperature to obtain reliable diffraction patterns.

## 3. Results and discussion

### 3.1. X-ray diffraction characterization of the thermally-induced activation

Macroscopically, hydrated GUT-2 is a crystalline, white powder. At an atomistic level, it consists of individual strands of a coordination polymer, connected by $H_2O$ molecules as illustrated in Fig. 1 (a)-(c). These $H_2O$ molecules form hydrogen bonds of the type $O\cdots H_w-O_w-H_w\cdots O$ (with atoms being part of the $H_2O$ molecule denoted by the subscript 'w') between the oxygen atom belonging to the carboxylate groups of the linkers. The individual strands of GUT-2 consist of $Zn^{2+}$ metal centres that are connected via 3-(2-methyl-1H-imidazole-1-yl)propanoate linkers. Each $Zn^{2+}$ ion is surrounded by two oxygen atoms from the carboxylate groups and two nitrogen atoms from the 2-methyl-imidazolate ring, forming a tetrahedral geometry with a coordination number of four. Overall, the crystal structure solution of hydrated GUT-2 that we obtained by single-crystal diffraction is fully consistent with the one proposed in Ref.[33]. This is shown by a direct comparison of the two sets of crystallographic data (see Table 1), which testifies to the reproducibility of the GUT-2 synthesis.

The activation of GUT-2 can be performed, for example, by heat treatment. The $H_2O$ molecules in hydrated GUT-2 efficiently leave the framework pores, for example, at a temperature of 50°C (with details on the kinetics of the process discussed in Section 2.3). This results in completely dehydrated but still intact, colorless, block-shaped single crystals. The activated form of GUT-2 (like the hydrated one) crystallizes in an orthorhombic crystal system, the space group changes from *Pcca* to *Pccn* and the primitive unit cell significantly contracts in two directions for the activated form (see Table 1).

Calculating the volumes of the unit cells listed in Table 1 yields a shrinkage by a factor of more than 2. At first glance, this appears like a drastic change; a closer inspection of the single crystal diffraction data, however, reveals that it is primarily a consequence of a reduction of the number of chemical formula units per unit cell by a factor of 2. This is the consequence of a minor rearrangement of the pores due to removal of the direct connections between neighboring strands as a consequence of the desorption of the $H_2O$ molecules. In turn the translational periodicity in the plane perpendicular to the polymer chains (i.e., in the *ac*-plane) changes: The unit cell of the hydrated form in Fig. 1 (b) contains four (partly blocked) channels shown as grey-shaded areas, which run in b-direction (i.e., in the direction parallel to the polymer chains). Conversely, in the activated form shown in Fig. 1 (e) only two (now completely open) channels are contained. Moreover, the unit cell in the *ac*-plane is rotated such that the a and c directions of the activated form run roughly parallel to the diagonals of the unit cell in the hydrated form and vice versa. Still, even when considering the factor of two arising from the modified translational periodicity, the volume of the unit cell of the activated form is reduced by around 3 %. This occurs despite the considerably higher temperature at which the unit cell of the activated form has been determined (100°C vs. -173°C for the fully hydrated form). The volume reduction occurs also despite the loss of H-bonding between polymer strands in *c*-direction and is, thus, primarily attributed to the $H_2O$ molecules not only strengthening the bonding between polymer strands but also serving as (weak) spacer units.

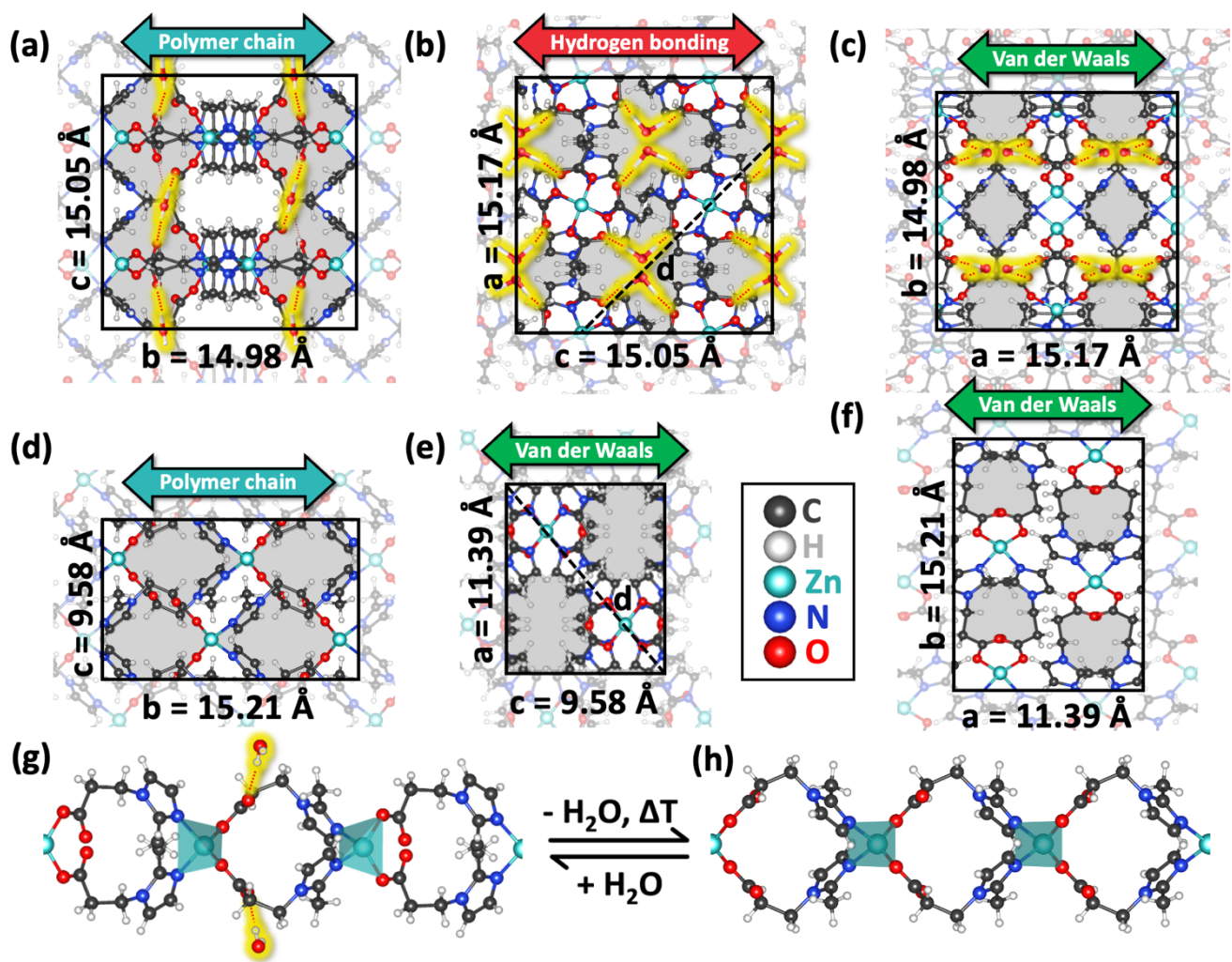

**Fig. 1** Framework structure within the orthorhombic crystal structures of hydrated (**a**)-(**c**) and activated GUT-2 (**d**)-(**f**) along all three unit cell axes. Hydrogen bonding in the hydrated form is indicated using red dotted lines (highlighted in yellow) connecting water molecules (also highlighted in yellow) to oxygen atoms in the carboxylates. Pores are shaded in grey. Double-sided arrows indicate the type of chemical bonding dominating in specific directions. Black, dashed lines indicate the diagonal *d* that will become relevant in the later discussion. Moreover, a single polymer strand of hydrated GUT-2 (**g**) and of its activated form (**h**) is projected along the axis connecting the centres of two adjacent pores. Transparent cyan polyhedra show the tetrahedral bonding situations around the $Zn^{2+}$ ions. The illustrations visualizing the crystal structures are generated using OLEX2 (version 1.5). [34]. Please note that for naming the crystallographic directions in the activated form we do not use the Niggli naming convention,[53] as discussed in detail in the caption of Table 1.

Additionally, the removal of $H_2O$ molecules results in slight changes of the linker alignments (see panel (g) and (h) of Fig. 1), resulting in a rotation of the 2-methyl-imidazolate linkers coordinated to the $Zn^{2+}$ ions. This also cages the methyl group orientation on the imidazole rings. The overall extent of these rearrangements is, however, rather minor, which indicates that the fundamental framework structure remains largely intact upon thermal activation. From a practical point of view, the most relevant difference between both forms of GUT-2 is that in the hydrated form the hydrogen-bonded $H_2O$ molecules block the pores running in the directions of the polymer strands (Fig. 1 (b)), which is no longer the case in the activated form (see Fig. 1 (e) and also Fig. S1 in the Supporting Information).

Despite this opening of additional channels in the activated form of GUT-2, the void space of the material calculated by the contact surface method[49] on the basis of the single crystal structure (as implemented in Mercury[48] (version: 2024.2.0)) hardly changes upon activation. It remains at a comparatively low value of roughly 2 %. This is in sharp contrast to the pronounced pore-opening transition observed, for example, in MIL-53 upon exposure to a variety of gases.[49,50] These different behaviors of GUT-2 and MIL-53 are not unexpected considering that here we are dealing with a system consisting of essentially 1D coordination polymer strands held together mostly by van der Waals forces and hydrogen bonds, while MIL-53 is a highly porous 3D-connected MOF. From the contact surface analysis[51] we can also conclude that the voids in GUT-2 can accommodate molecules with a maximum probe radius of 1.2 Å. When choosing larger probe sizes, we do not observe any detectable voids. This is insofar relevant, as it indicates that both forms of GUT-2 lack sufficient space to host additional $H_2O$ molecules, which are typically associated with a probe radius of 1.4 Å.[52]

**Table 1.** Structural parameters for the hydrated and activated form of GUT-2 according to single crystal X-ray diffraction measurements. Please note that in naming the different crystallographic directions for the activated form, we do not follow the Niggli naming convention[46] to allow a more direct comparison to the structural parameters (including thermal expansion coefficients) of the hydrated form. The cell parameters (*a*, *b* and *c*) following Niggli convention are, thus, provided in brackets. The parameter Z denotes the number of (chemical) formula units within the unit cell. For the hydrated form, the single crystal structure solution exhibits a residual value ($R_1$) of 2 % (for definition of $R_1$ see Chapter S2 in the Supporting Information) when considering reflections for which the intensity satisfies the condition I $\geq 2\sigma$. This means that only reflections where the measured intensity exceeds twice its estimated standard deviation are included in the calculation. In contrast, the activated form of GUT-2 at 100°C yields a higher $R_1$ value of 8 %, which can be attributed to the elevated temperature at which the measurement has been performed. For a more in-depth discussion of this topic, the reader is referred to Chapter 2 in the Supporting Information (especially Fig. S2), where also a more extensive version of this table is provided.

|  | Hydrated GUT-2 (-173°C)[33] | (1) Hydrated GUT-2 (-173°C) | (2) Activated GUT-2 (100°C) |
|---|---|---|---|
| Formula | $C_{14}H_{18}N_4O_4Zn \cdot H_2O$ | $C_{14}H_{18}N_4O_4Zn \cdot H_2O$ | $C_{14}H_{18}N_4O_4Zn$ |
| Weight [g/mol] | 389.73 | 389.71 | 371.69 |
| Temperature [K] | 100 | 100 | 373 |
| a [Å] | 15.1861(13) | 15.1721(3) | 11.3850(7) [b] |
| b [Å] | 15.0082(13) | 14.9839(3) | 15.2053(7) [c] |
| c [Å] | 15.0568(13) | 15.0445(3) | 9.5687(6) [a] |
| $\alpha = \beta = \gamma$ [°] | 90 | 90 | 90 |
| Volume [Å$^3$] | 3431.7(5) | 3420.17(12) | 1656.46(16) |
| Z | 8 | 8 | 4 |
| Crystal system | Orthorhombic | Orthorhombic | Orthorhombic |
| Space group | Pcca | Pcca | Pccn |
| Crystal size [mm$^3$] | 0.05 × 0.05 × 0.04 | 0.17 × 0.12 × 0.09 | 0.17 × 0.12 × 0.09 |
| $R_1$, $wR_2$ (I $\geq 2\sigma$) | 0.0212, 0.0533 | 0.0344, 0.0785 | 0.0839, 0.2335 |

As a further validation of the structures determined by single crystal X-ray diffraction, we also performed geometry optimizations using state-of-the-art dispersion-corrected density functional theory. As detailed in the Supporting Information, these simulations yield only very minor changes of the lattice constants. These minor deviations are at least in part caused by the fact that the simulations are performed at essentially 0 K. Also, the positions of the atoms within the unit cell remain virtually the same indicating that the proposed structures represent stationary points of the potential energy surface of GUT-2.

Additionally, we verified the mechanical stability of both GUT-2 structures by testing the Born stability criterion,[53] which is based on the elastic tensor elements of the structures. In passing we note that in a recent work, some of us managed to extract elements of the elastic tensor of the hydrated GUT-2 structure, via a combination of Brillouin light scattering spectroscopy and theoretical methods, yielding excellent agreement between experiment and theory.[54] Using the results from that study and calculating the elastic constants also for activated GUT-2, one can show that the Born stability criterion[53] is fulfilled for both structural solutions from Table 1. This testifies to their mechanical stability (see Chapter S10).

The following experiments were carried out to determine the thermal expansion of GUT-2 in its hydrated and activated forms and to analyze its activation and hydration kinetics. They were performed on isotropic powder samples. This is motivated by the fact that powder diffraction experiments provide phase-averaged structural information, minimizing the impact of local inhomogeneities or domain effects.[55]

Before discussing thermal expansion, a crucial aspect needs to be addressed that will be central to our analysis of the anisotropic thermal expansion coefficients: the different types of chemical bonding types present in GUT-2. The hydrated form of GUT-2 features three distinct types of interactions, as highlighted in Fig. 1 using double-sided arrows: (i) covalently linked polymeric chains extending along the *b*-direction, (ii) a network of hydrogen bonds along the *c*-direction that link polymer chains into 2D sheets, and (iii) van der Waals interactions along the *a*-direction. The latter represent the weakest type of bonding, but still, they significantly contribute to the overall cohesion of the structure. Upon activation, where $H_2O$ molecules are removed, the polymer chain backbone remains oriented along the *b*-direction, maintaining the primary structural integrity of the MOF. However, due to the absence of $H_2O$ molecules, hydrogen bonds cease to exist. As a result, a structural rearrangement occurs, such that after activation polymer chains are held together primarily by van der Waals interactions in all directions perpendicular to the *b*-direction. Importantly, these van der Waals interactions do not remain identical to those in the hydrated form, as the framework adjusts to compensate for the loss of hydrogen bonding, thereby altering the overall bonding environment (see above discussion).

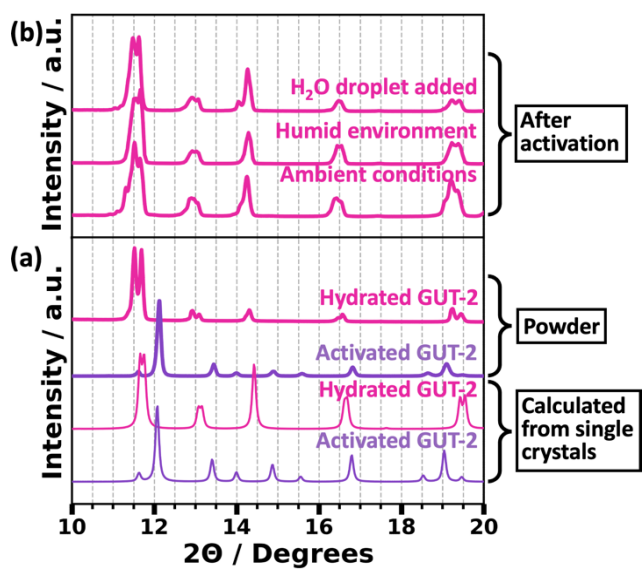

**Fig. 2** Powder X-ray diffraction (PXRD) patterns of GUT-2 in various stages of activation. PXRD patterns shown by thick lines are experimentally measured curves, while thinner curves correspond to simulated data. In panel (**a**), diffractograms of activated GUT-2 (purple curve) and hydrated GUT-2 (pink curve) determined calculated based on the single crystal diffraction data are compared to powder patterns of hydrated GUT-2 measured right after the synthesis and after activation by thermal treatment. Additionally, panel (**b**) contains the diffractograms for the activated powder left at ambient conditions (relative humidity of 30 % and a temperature of 25°C) for two weeks, for the powder stored in humid environment (relative humidity of 100 %) for two days, and for the powder after direct treatment with a drop of water.

To interpret the powder data, we calculated the expected powder diffractograms of hydrated and activated GUT-2 based on the single crystal results using Mercury[48] (version: 2024.2.0). They are shown in Fig. 2 (a) as thin pink and purple lines, respectively. Their peak positions agree very well with the diffractogram of the studied powder (thick pink curve) achieved by grinding the as-synthesized material using a porcelain mortar (see Method section) and a powder achieved by activating the MOF in vacuum at 100°C for 12 hours (thick purple curve). Minor differences in peak positions are attributed to the different temperatures at which the structures were measured: the single crystal data for the fully hydrated form of GUT 2 were determined at -173°C and those for the activated form at 100°C (see above), while the powder patterns were measured at room temperature. Interestingly, after activation the dehydrated GUT-2 powder can be rehydrated in a variety of ways. This is illustrated in Fig. 2 (b) for the case of leaving the sample for two weeks at typical lab conditions (~30 % humidity), for the case of keeping the sample in a sealed vessel containing $H_2O$ at the base (100 % humidity) and for putting a drop of $H_2O$ on top of the sample, which causes an instantaneous conversion back to the hydrated form. Overall, the diffractograms for all these cases agree very well with that of the as synthesized form. Minor deviations, for example, for the diffractogram of the sample rehydrated over two weeks at ambient conditions suggest a certain level of defects introduced by extended (de)hydration cycles.

### 3.2. Varying sample temperature and hydration state

The ability of GUT-2 to undergo a reversible reaction to switch between its hydrated and activated crystal structure makes it an ideal system to study the impact of the presence of guest molecules on the thermal expansion of MOFs. To determine the anisotropic thermal expansion coefficients of GUT-2, temperature-dependent PXRD measurements were performed. The temperature curves for the sample treatment are shown in Fig. 3 (a). Starting from the hydrated form at 25°C, the experiment followed a sequence of two sets of cooling and heating cycles (C1 and H1 as well as C2 and H2). Between the two sets of cycles the originally (partially) hydrated GUT-2 was thermally activated by keeping the sample at 100°C for 12 hours. This yielded diffractograms for one cooling and one heating cycle for each form (hydrated and activated). The lowest temperature reached in C1 was -190°C. The highest temperature reached during H2 was 250°C, a temperature that is just below the decomposition temperature of GUT-2, which has been reported to be 270°C.[33]

The most prominent diffraction features of the hydrated form are clearly resolved (and indexed) during the cycles C1 and H1. Comparing the temperature-dependent diffraction patterns in Fig. 3 (b) and (c) with (d) and (e) reveals that all peaks observed after GUT-2 activation are also present (with lower intensities) before the annealing at 100°C. Notably, the most prominent peak associated with the activated form (the (110) peak at 12.1 degrees) as well as the peak at 13.4 degrees (the (111) peak of the activated form) intensify with heating during process step H1. In contrast, during the initial cooling C1 only a very minor change in the relative peak intensities is observed. This suggests that partial activation of the powder has occurred already during the preliminary test-runs prior to the actual C1 measurements and this activation continues during the heating in processing step H1 (for further details see Fig. S10 in the Supporting Information).

After the extended heat treatment (prior to processing steps C2 and H2) only diffraction features of the activated form prevail. To verify that the aforementioned pre-activation of the MOF sample has indeed occurred and to comprehensively analyze all structural data obtainable from the diffractograms, Rietveld refinements are performed. A comparison of the Rietveld refinements of the C2 and H1 measurements at a temperature of -150°C – the lowest common temperature – is shown in Fig. 3. Rietveld refinements at all other temperature set points are given in Fig. S4-S9 for all four temperature curves, while Tables S11-S14 list the statistical parameters assessing the quality of the fits. There, also a more in-depth discussion of the quality of the fits and the mathematical definition of the statistical parameters can be found.

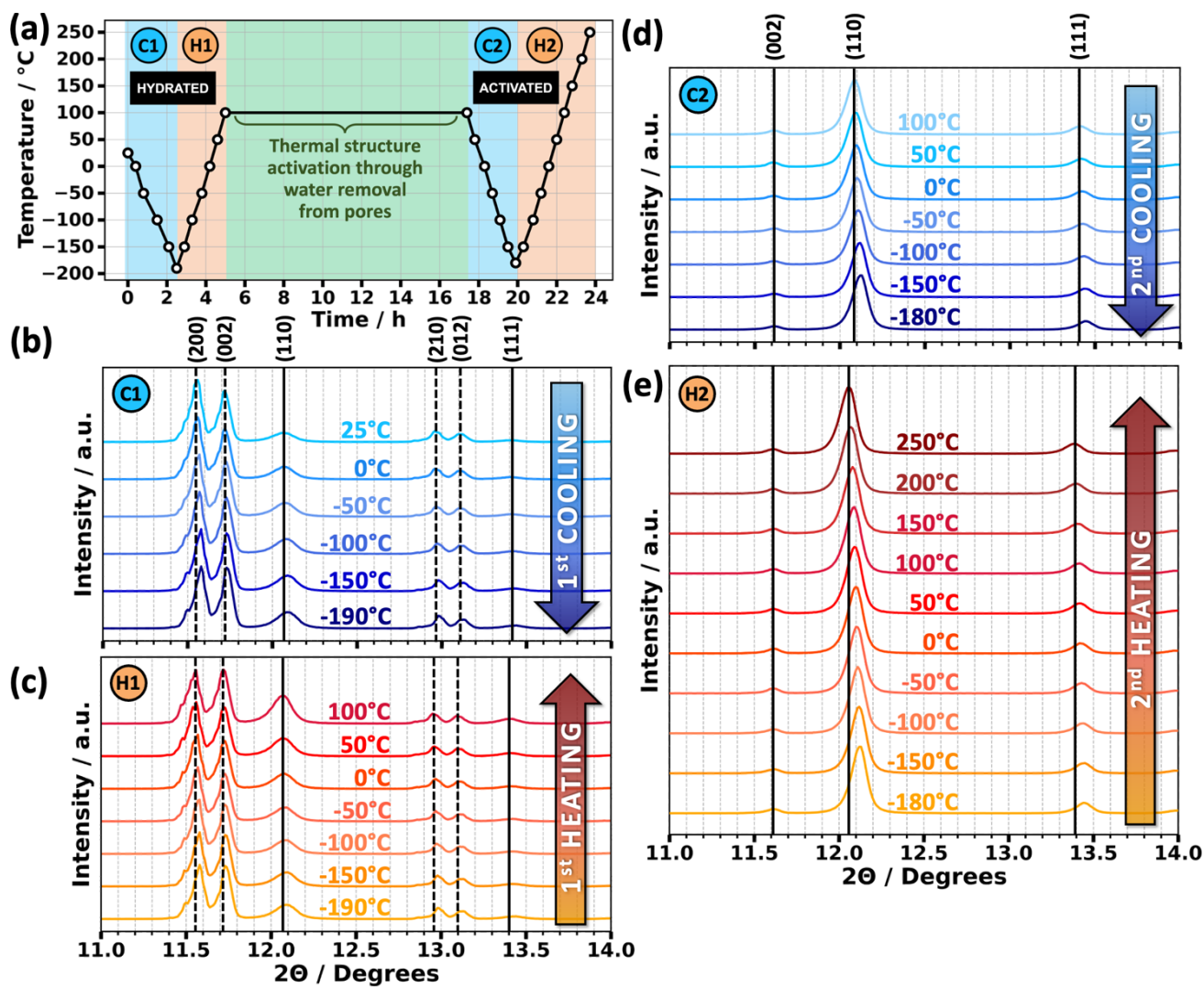

**Fig. 3** (**a**) Temperature sequence for powder X-ray diffraction study of hydrated and activated GUT-2 at fixed set points (black circles). The first cooling curve (C1) for the hydrated form starts at room temperature (25°C) and cools down to a temperature of -190°C. This is followed by the first heating curve (H1) where, the sample is heated to up to 100°C. Subsequently, the sample is held at this this temperature for 12 h to fully activate GUT-2 by dehydrating the pores. Subsequently, the activated sample undergoes a second cooling cycle (C2) down to -180°C, followed by a final heating cycle (H2) up to 250°C. (**b**)-(**e**) show temperature-dependent diffraction pattern in the 2Θ range between 11 degrees and 14 degrees obtained from both heating and cooling curves. Peaks associated with hydrated GUT-2 are highlighted by dashed vertical lines, whereas solid vertical lines indicate the peaks of the activated form. These lines represent the peak position at the highest temperature. All these experiments were performed in vacuum.

In short, in the context of Rietveld refinement, the parameter $R_p$, or profile residual, serves as a metric for assessing the agreement between experimental and simulated diffraction data.[56] It is defined as the sum of the absolute differences between observed and calculated intensities, divided by the total observed intensity. While a lower $R_p$ value suggests a better fit, it should be interpreted with caution, as it does not consider factors such as background noise or preferred orientations. The most straightforward refinement can be done for the diffractogram shown in Fig. 4 (c), where the crystal structure of activated GUT-2 serves as an ideally suited refinement model. The $R_p$ value for this refinement amounted to 17.4 %. Compared to typical $R_p$ values for well-ordered crystalline materials, this might appear rather large, but in the present case It is most likely just caused by the deviation from a perfectly isotropic arrangement of the crystallites. The Rietveld refinement of the H1/C1 diffraction pattern using only the hydrated model (Fig. 4 (a)) provides an incomplete description of the data, leaving key experimental peaks unexplained. Incorporating the activated form of GUT-2 as a secondary phase significantly enhances the fit (Fig. 4 (b) and Fig. S5 and S7).

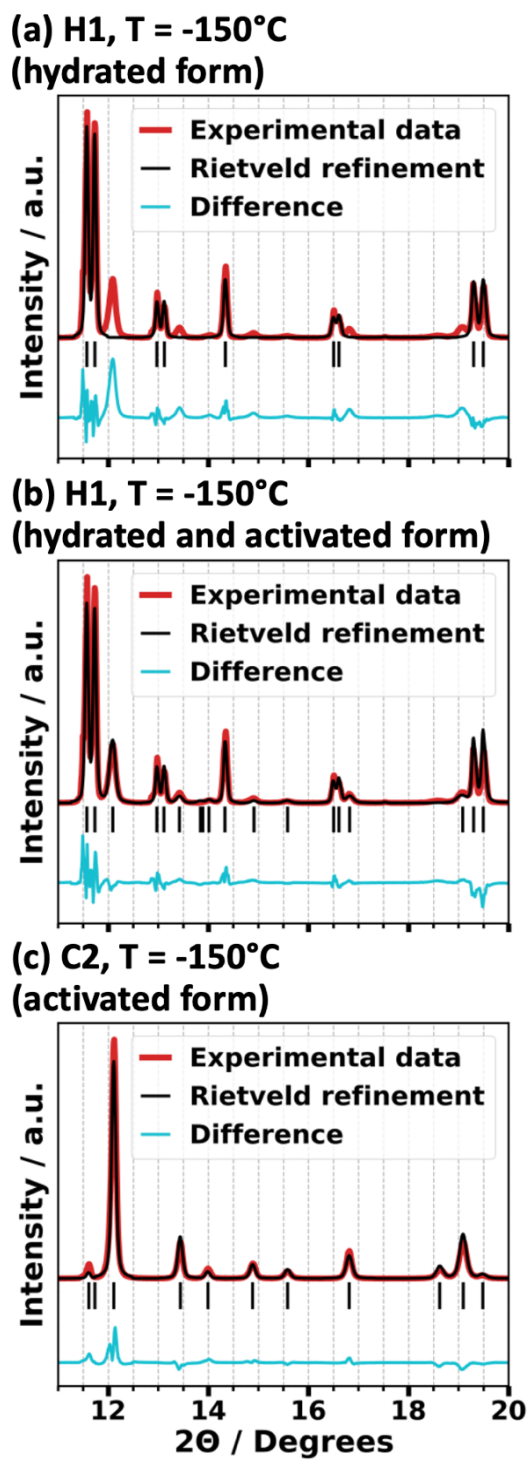

**Fig. 4.** Overlay of the powder X-ray diffraction patterns (red curves) measured at -150°C and Rietveld refinements (black curves). Panel (**a**) corresponds to the first heating curve (H1) using only the hydrated form of GUT-2 as a model, while panel (**b**) includes both the activated and hydrated forms of GUT-2. The Rietveld refinement of the second cooling curve (C2) shown in panel (**c**) was performed using only the structure of activated GUT-2 as model. The calculated Bragg peaks are shown as vertical lines. The difference between the experimental data and the Rietveld refinement is shown as a blue curve below the actual data.

Phase quantification shows that during each cooling cycle (prior to full activation), approximately 8 % of the sample are activated in the temperature range from 25°C to -50°C over a timespan of 2 hours, with no additional activation occurring at lower temperatures (see also Fig. S10 and associated discussion). During the first heating curve an additional activation of about 10 % occurs between -50°C and 100°C within 24 minutes. The details of the corresponding phase quantification are provided in Fig. S11 in the Supporting Information. This analysis confirms that in vacuum during the first stages of the initial cooling and during the first heating a gradual activation of GUT-2 occurs, which is then completed during the 12h annealing at 100°C.

Information about the crystallographic structure in terms of coherent crystallite size and the presence of microstrains are accessible via the widths of the diffraction peaks. To quantify the influence of instrumental peak broadening on our diffraction measurements, the PXRD pattern of the standard material $LaB_6$ was recorded under settings identical to those described in the Method Section 3.3, with measurements performed at room temperature. Extrapolation of the $LaB_6$ full widths at half maximum (FWHM) in the 11.6 degrees to 12.1 degrees range yielded an average of around 0.03 degrees (see Fig. S3 of the Supporting Information for more details), which is minimal. Therefore, no correction for instrumental broadening is needed when calculating the crystallite size. What is critical, however, is that the FWHM values are Rachinger corrected which is an iterative numerical technique that allows one to deconvolute the overlapping contributions of the $K\alpha_1$ and $K\alpha_2$ lines.[57]

Table 2 summarizes the thus obtained crystallite sizes of GUT-2 using the Scherrer equation[58] for both cooling and heating cycles. For the hydrated form the crystallite sizes (determined from the (200) reflection in the C1 and H1 series) are comparably large (111 ± 4 nm and 111 ± 3 nm, respectively). This demonstrates the good structural integrity of GUT-2 when $H_2O$ molecules are present in the pores. In contrast, the activated form (analyzed based on the (110) reflection) shows significantly smaller crystallite sizes of 63 ± 1 nm and 69 ± 5 nm in the C1 and H1 series. Interestingly, in the second measurement set (C2 and H2), the activated form exhibits larger domains (76 ± 2 nm and 77 ± 1 nm), indicative of a structural annealing during the extended heat treatment.

**Table 2.** Estimated sizes of hydrated and activated GUT-2 crystals using the Scherrer equation [58] for both cooling (C1 and C2) and heating cycles (H1 and H2).

| Measurement series | Laue indices | Crystal domain | Crystallite size [nm] |
|---|---|---|---|
| C1 | (200) | Hydrated | 111 ± 4 |
|  | (110) | Activated | 63 ± 1 |
| H1 | (200) | Hydrated | 111 ± 3 |
|  | (110) | Activated | 69 ± 5 |
| C2 | (110) | Activated | 76 ± 2 |
| H2 | (110) | Activated | 77 ± 1 |

### 3.3. Quantifying the thermal expansion and its dependence on the hydration state

The Rietveld refinements discussed in the previous section are valuable not only for phase quantifications but also for determining the anisotropic thermal expansion coefficients. Using the temperature-dependent cell parameters obtained from these refinements (shown in Fig. 4), the thermal and volumetric expansion coefficients can be calculated through the following equations:

$$\alpha_L = \left(\frac{\partial L}{\partial T}\right)_p \frac{1}{L_0} \quad \text{with } L = \{a, b, c, d\}$$

$$\alpha_V = \left(\frac{\partial V}{\partial T}\right)_p \frac{1}{V_0}$$

In the first equation, $\alpha_L$ is the linear thermal expansion coefficient, defined as the relative rate of change with temperature $T$ of one of the three unit cell lengths in the crystallographic directions $a$, $b$ and $c$. For reasons that will become apparent below, also the relative changes in directions parallel to the diagonals of the unit cells, $d$, were evaluated. As the unit cells in the two forms of GUT-2 are rotated relative to each other in the $ab$-plane (see Fig. 1 (b) and (e)), the diagonals in one form essentially correspond to the unit cell directions in the other. The second of the above equations introduces the volumetric thermal expansion coefficient, $\alpha_V$, which quantifies the fractional volume change with temperature. The linear thermal expansion coefficients for GUT-2 as well as the volumetric expansion coefficient are obtained from linear fits of the temperature-dependent cell parameters, as shown in a side-by-side comparison of the hydrated and activated forms in Fig. 5. $L_0$ and $V_0$ are set to the values of the temperature at 0°C. Conceptually, also non-linear terms are expected to contribute as there is no reason to assume that the thermal expansion of GUT-2 remains exactly the same for all studied temperatures. However, the linear fits represent the experimental data rather well and the uncertainty of the individual datapoints makes fitting higher-order polynomials futile. Thus, we restrict the analysis to the linear expansion coefficients summarized in Table 2 for all cooling and heating processes.

Hydrated and activated GUT-2 show generally low volumetric thermal expansion values of around 19×10$^{-6}$ K$^{-1}$ to 26×10$^{-6}$ K$^{-1}$.

Notably, the temperature-induced volume change shows a weak hysteresis effect, especially during the cooling cycles. In its hydrated form, GUT-2 exhibits nearly ZTE along the $b$-axis, with expansion coefficients ranging from approximately 1.2×10$^{-6}$ K$^{-1}$ to 3.7×10$^{-6}$ K$^{-1}$ (compare Table 3). This direction is characterized by strong covalent bonding within the polymer chains, which significantly restricts thermal expansion due to the high bond strength and the narrow, apparently almost symmetric potential energy well associated with changing the length of $b$. A more quantitative discussion of how strong bonds typically lead to a reduced thermal expansion is provided in Chapter S8 in the Supporting Information. Concerning thermal expansion in the direction of the $c$-axis of the hydrated form, where hydrogen bonds connect the individual polymer strands, the expansion coefficient increases to around 5.9×10$^{-6}$ K$^{-1}$ to 6.9×10$^{-6}$ K$^{-1}$ reflecting the fact that hydrogen bonds are significantly weaker (and apparently more anharmonic) than covalent bonds. Finally, the $a$-axis, in which inter-chain interactions are predominantly governed by van der Waals interactions, exhibits the highest thermal expansion coefficient, reaching values of 8.0×10$^{-6}$ K$^{-1}$ to 8.9×10$^{-6}$ K$^{-1}$. This trend is again in line with expectations, as van der Waals forces are significantly weaker than hydrogen or covalent bonds and are particularly anharmonic (considering, e.g., Lenard-Jones potentials; see also discussion in Chapter S8). This allows the framework to deform more easily upon heating.

Upon activation of the framework, the $b$-direction – which for our naming convention again corresponds to the polymer chain direction – continues to exhibit essentially ZTE, with values remaining close to 1.8×10$^{-6}$ K$^{-1}$. Regarding thermal expansion in directions perpendicular to the chain, it is not useful to directly compare $\alpha_a$ and $\alpha_c$ between both systems due to the rotated unit cell. Instead, it is sensible to analyze the relative change of the Zn-Zn distance mentioned above, which (as illustrated in Fig. 1 (e)) corresponds to the relative change of the length of the diagonals of the unit cells in the $ac$-plane, $\alpha_d$. Notably, due to the orthorhombic space group the expansion coefficients for both diagonals are identical in the activated form. Interestingly, $\alpha_d$ of activated GUT-2 is clearly larger than $\alpha_c$ of the hydrated form and is in the same range as the value for the van der Waals bonding direction in the hydrated form, $\alpha_a$. This is consistent with the absence of hydrogen bonds in activated GUT-2.

**Table 3.** Anisotropic lattice and volumetric thermal expansion coefficients for the cooling (C1, C2) and heating curves (H1, H2) obtained via linear fits through the cell parameters that were determined via Rietveld refinements. While the data for C1 and H1 are characteristic of the hydrated form of GUT-2, C2 and H2 correspond to the fully activated GUT-2. $a$, $b$, and $c$ refer to the lattice constants, $d$ to the diagonal (see Figure 1 (b) and (e)) and $V$ to the volume of the unit cell.

|   |    | $\alpha_a$ [10$^{-6}$ K$^{-1}$] | $\alpha_b$ [10$^{-6}$ K$^{-1}$] | $\alpha_c$ [10$^{-6}$ K$^{-1}$] | $\alpha_d$ [10$^{-6}$ K$^{-1}$] | $\alpha_V$ [10$^{-6}$ K$^{-1}$] |
|---|----|------|------|------|------|------|
| Hydrated form | C1 | 8.94 | 1.24 | 5.87 | 7.41 | 20.0 |
|               | H1 | 8.01 | 3.74 | 6.88 | 7.45 | 18.8 |
| Activated form | C2 | 5.79 | 1.90 | 13.0 | 8.77 | 19.6 |
|                | H2 | 7.53 | 1.76 | 16.3 | 11.8 | 25.5 |

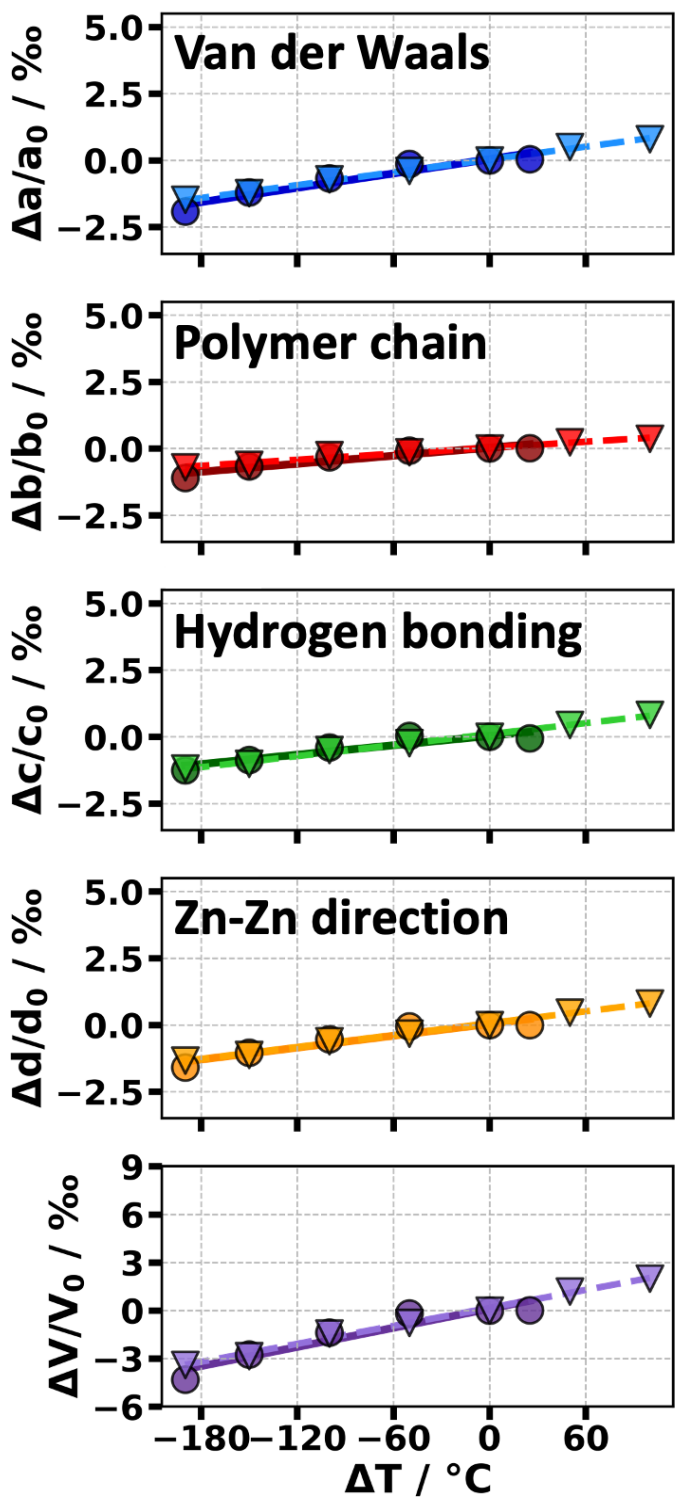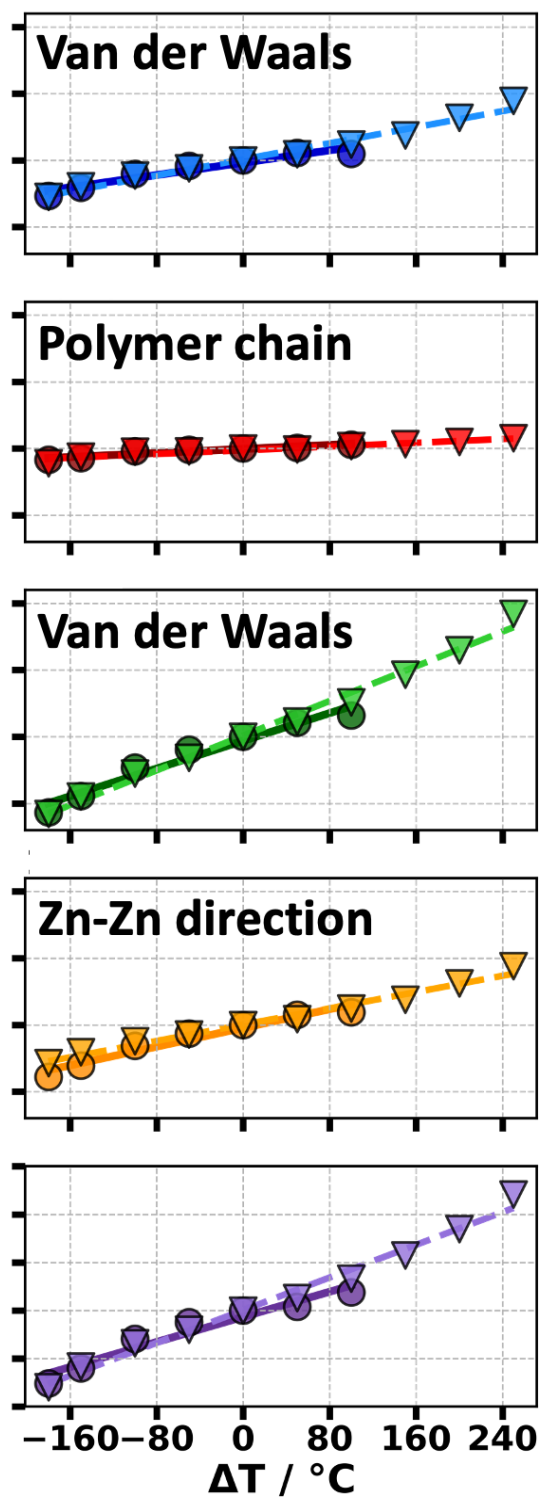

**Fig. 5** Temperature-induced changes of the three cell parameters (*a*, *b* and *c*) as well as of the diagonal (*d*) and the cell volume (*V*) relative to the values measured at a temperature (*T*) of 0°C. The data represent the situation of hydrated GUT-2 (**a**) and activated GUT-2 (**b**). The temperature change is given relative to 0°C and the unit-cell parameters have been extracted based on Rietveld refinements. Data points represented by circles (O) correspond to cooling runs, while those shown as inverted triangles (∇) belong to heating runs. Solid lines indicate the linear fits for the cooling processes and dashed lines represent the fits for the heating processes.

### 3.4. Crystallization kinetics

A second set of experiments was performed to assess the possibility of controlling the dehydration and rehydration process of GUT-2 under standard laboratory conditions. To probe these aspects, separate temperature-dependent PXRD measurements were carried out in air (relative humidity of 30 %) rather than in vacuum (for details on the experimental setup see Method Section 3.4). Starting with the hydrated as-synthesized GUT-2 powder sample – which is identical to the sample used in previous experiments – the material was activated at a constant temperature of 50°C and PXRD patterns were recorded every 40 minutes (reported in the first column of Fig. S12 in the Supporting Information). Rietveld refinements were employed to quantify this phase transformation, allowing for a quantitative tracking of the hydration state of GUT-2 as a function of time. The resulting phase fractions are displayed in Fig. 6 (a). They reveal that full activation of the MOF under these conditions requires approximately 6 hours. A second experiment (Fig. S12 (a)) showed that when the sample is heated to 90°C (again in air), full activation occurs within 30 minutes.

After activation, the dehydrated GUT-2 sample was left in air at 30 % humidity and at room temperature (25°C) to explore the rehydration process. Again, PXRD measurements captured the structural changes as $H_2O$ gradually diffused back into the framework. The corresponding phase quantifications, summarized in Fig. 6 (b), show that $H_2O$ adsorption under these conditions required nearly 2 days to reach completion.

To gain deeper insights into the mechanisms regarding $H_2O$ uptake and release in GUT-2, we employed the Avrami equation [59–61]. It is based on a widely used phenomenological model [62–64] that describes the kinetics of isothermal phase transitions, particularly providing insights into geometrical evolutions of the crystals during the transformation process. Moreover, the model was applied to study the structural evolution of the MOFs ZIF-8[65] and ZIF-67[66].

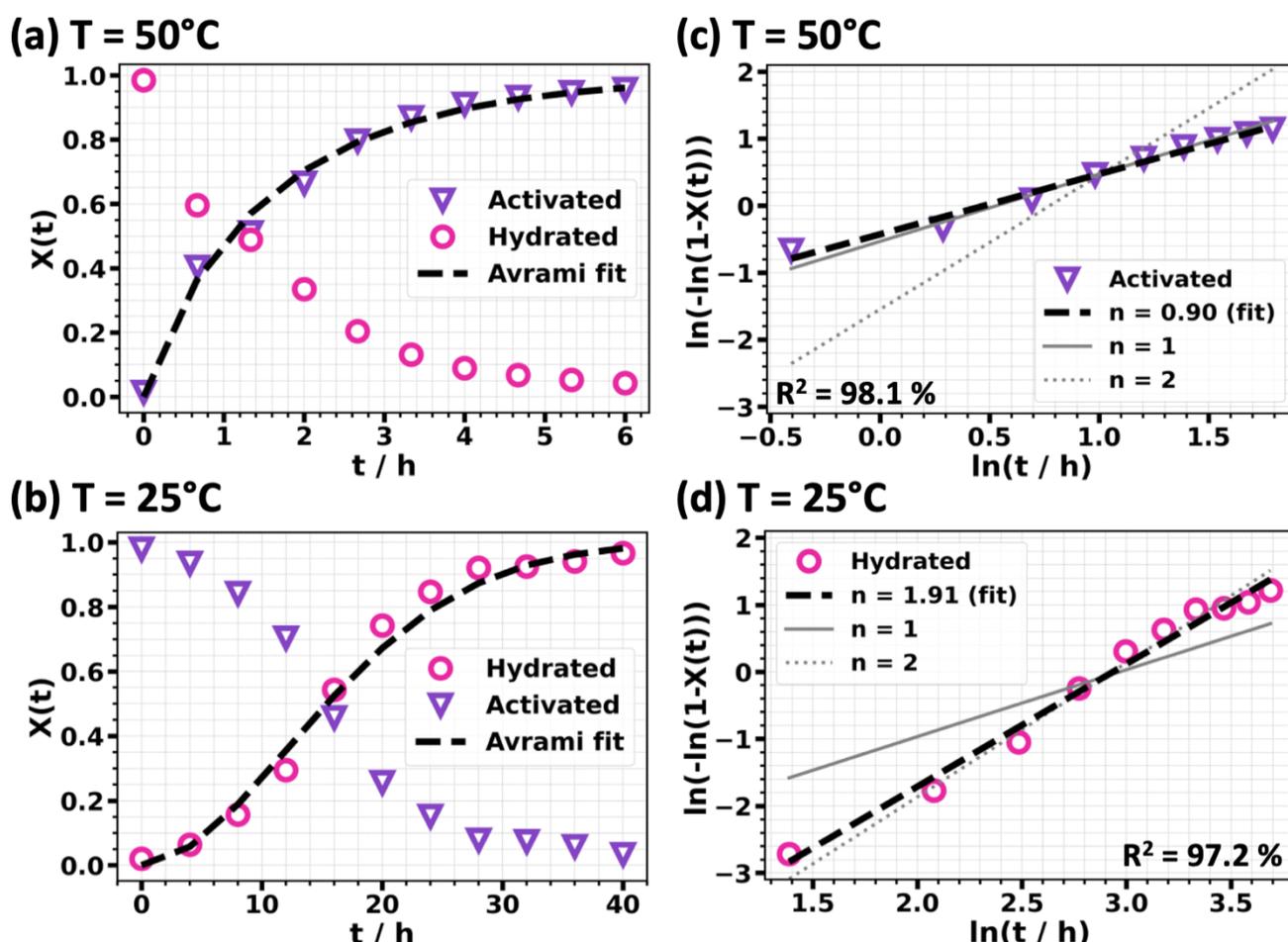

**Fig. 6** Evolution of the hydration state of GUT-2 powder during the activation process at 50°C (**a**) and the corresponding re-hydration process at 25°C (**b**). The left panels include the resulting Avrami fit (dashed black lines) based on the extracted constants from the linearized form of the Avrami plots shown on the right-hand side. *X(t)* denotes the fraction of the compound that has been converted at a certain time t, ranging from 0 (untransformed) to 1 (fully transformed). Data points with purple, inverted triangles (∇) belong to the activated GUT-2 phase that forms at a constant temperature of 50°C, while data points illustrated with pink circles (O) belong to the hydrated GUT-2 phase that forms at a temperature of 25°C. For the linear fits the coefficient of determination ($R^2$) is given. The solid and dashed grey lines in panels (**c**) and (**d**) represent slopes of 1 and 2, respectively, and are shown for the sake of comparison.

The Avrami equation can be written as

$$X(t) = 1 - \exp(-kt^n)$$

with *X(t)* being the transformed volume fraction as a function of the time *t*, *n* corresponding to the Avrami exponent and *k* representing the reaction rate constant. The Avrami exponent *n* has been related to the dimensionality of crystal growth[67] in particular if it assumes integer numbers.[64] While the classical formulation of the Avrami equation does not account for diffusion-controlled processes (as they occur for GUT-2 hydration and activation) the Avrami formalism still serves as an effective tool to approximate and interpret the overall transformation kinetics. In such a case, associating integer values of *n* with the dimensionality of the process might not always be fully justified. Nevertheless, changes in *n* can be interpreted as a clear indication for changes in the underlying mechanism for the growth/shrinkage of specific structures (here, during the adsorption/desorption of $H_2O$ molecules).

To determine the Avrami exponent for the activation and re-hydration of GUT-2, a linearized form of the Avrami equation is used. By plotting *ln[-ln(1-X(t))]* versus *ln(t)*, the exponent *n* can be extracted directly from the slope of the resulting straight line. For the activation of GUT-2 at 50°C, the Avrami exponent is determined to be 0.90, i.e., slightly below 1, which would typically be interpreted as a low-dimensional process. For the rehydration of GUT-2 the exponent extracted from the fit amounts to 1.91, indicative of a mechanism with larger dimensionality than the activation process. We note that the difference in the Avrami exponents is observed for processes at different temperatures, which can have an impact on the dimensionality of the growth of adsorption/resorption sites. Moreover, the diffusion along the polymer axis (in *b*-direction; Fig. 1 (b)) is hindered by adsorbed $H_2O$ molecules, as they block the channels. In part this is also true for water diffusion along the *a*-axis (see Fig. 1 (a)). This can influence activation and hydration differently, as the activation process will typically start from the surface of the crystallites and then proceeds towards their interior; upon hydration, the $H_2O$ molecules will again first adsorb and block the channels close to the surface, but now this is expected to delay the further conversion in the interior of the crystallites as at least along two directions water-diffusion is hindered. These considerations show that it is realistic that the kinetics of hydration and activation of GUT-2 are qualitatively different (as indicated by different Avrami exponents). To what extent the dimensionality of the processes plays a role, remains, however, elusive.

## 4. Conclusions

In conclusion, we show that the coordination polymer GUT-2, consisting of $Zn^{2+}$ centres connected by 3-(2-methyl-1H-imidazol-1-yl)propanoate linkers, can be reversibly hydrated and activated. In the hydrated form, $H_2O$ molecules generate bridging hydrogen bonds between neighbouring polymer strands and at the same time block the pores of GUT-2 in the direction parallel to the polymer chains. Activation of GUT-2, besides breaking the hydrogen bonds between polymer chains, also triggers a modification of the pore structure such that the crystallographic unit cell is rotated in the plane perpendicular to the polymer chains. Additionally, its cell volume is reduced by more than a factor of 2 due to the concomitantly reduced number of atoms in the unit cell. The structural properties of GUT-2 are initially determined by single-crystal diffraction and subsequently confirmed by Rietveld refinement of powder data.

In the current study, GUT-2 is activated by heating crystallites either in air or in vacuum. In air and at 50°C, full activation takes approximately six hours, while rehydration under ambient conditions and 30 % relative humidity takes roughly two days. Notably, the timescales depend on the chosen conversion conditions (like the relative humidity of the atmosphere, the base pressure, or the sample temperature). An essentially instantaneous rehydration of the crystallites can be achieved by exposing them to a drop of liquid $H_2O$, where X-ray diffraction experiments suggest that the said rehydration processes induce a certain number of defects, but do not destroy the structural integrity of the studied materials.

Temperature-dependent X-ray diffraction experiments on suitably preconditioned samples in combination with Rietveld refinements allow the determination of the anisotropic thermal expansion coefficients of both forms of GUT-2. The results reveal an interesting correlation between the nature of the bonding interactions in a specific direction and the thermal expansion coefficient in that direction: Along the polymer chains, where covalent bonds dominates, particularly small thermal expansion coefficients typically around $2 \times 10^{-6}$ $K^{-1}$ are observed. Thermal expansion increases by a factor of around 3 to 4 in the less strongly bonded direction dominated by hydrogen bonds (in hydrated GUT-2) and increases even somewhat further in the Van der Waals bonded directions (in both forms of GUT-2). This suggests a correlation between the depths of the bonding potentials in the different directions and their degree of anharmonicity.

Finally, the Avrami equation is used to analyse the dynamics of activation and rehydration processes of GUT-2 and to gain further insights into the crystal formation. Interestingly, we find distinctly different Avrami exponents for the kinetics of the activation and hydration processes, which suggests fundamental differences between them. This is associated to the blocking of pores in polymer direction by adsorbed $H_2O$ molecules. Especially for the rehydration, this partially blocks the access of water molecules to the interior of the crystallites. Our findings demonstrate that minor changes in the hydration state of a coordination polymer or MOF can lead to significant functional adaptations in the properties of the studied materials. The results provide a compelling case for the strategic use of guest molecules to modulate the properties of porous frameworks.

## Author contributions

Conceptualization: E.Z., R.R., C.S., N.S.
Formal analysis: N.S.
Funding acquisition: E.Z, R.R.
Investigation: N.S., B.S., S.J., B.K., A.T., F.P.L.
Resources: E.Z, R.R., C.S.
Software: B.B., A.T., N.S., F.P.L.
Supervision: E.Z, R.R.
Visualization: N.S., F.P.L.
Writing – original draft: N.S
Writing – review & editing: E.Z, R.R.

## Conflicts of interest

There are no conflicts to declare.

## Data availability

The data generated in this work will be made available at the TU Graz Repository upon acceptance by a peer-reviewed journal: https://doi.org/10.3217/tst5n-0d276.

## Acknowledgements

This research was funded in part by the Graz University of Technology through the Lead Project Porous Material @ Work for Sustainability (LP-03). This research was primarily funded by the Austrian Science Fund (FWF) [10.55776/P34463]. For the purpose of open access, the author has applied a CC-BY public copyright license to any Author Accepted Manuscript version arising from this submission. The computational results have been obtained using the Vienna Scientific Cluster.

Supporting Information for

# Influence of pore-confined water on the thermal expansion of a zinc-based metal-organic framework

Nina Strasser,[a] Benedikt Schrode,[b] Ana Torvisco,[c] Sanjay John,[a] Birgit Kunert,[a] Brigitte Bitschnau,[d] Florian Patrick Lindner,[a] Christian Slugovc,[e] Egbert Zojer*[a] and Roland Resel*[a]

[a] Institute of Solid State Physics, NAWI Graz, Graz University of Technology, Petersgasse 16, 8010 Graz, Austria
[b] Anton Paar GmbH, Anton-Paar-Straße 20, 8054 Graz, Austria
[c] Institute of Inorganic Chemistry, NAWI Graz, Graz University of Technology, Stremayrgasse 9, 8010 Graz, Austria
[d] Institute of Physical and Theoretical Chemistry, NAWI Graz, Graz University of Technology, Stremayrgasse 9, 8010 Graz, Austria
[e] Institute of Chemistry and Technology of Materials, NAWI Graz, Graz University of Technology, Stremayrgasse 9, 8010 Graz, Austria

* Correspondence: egbert.zojer@tugraz.at, roland.resel@tugraz.at

## S1. Pores of hydrated and activated GUT-2

The influence of H$_2$O molecules on the pore structure of hydrated GUT-2 is visualized using a space-filling model, as presented in Fig. S1. By employing a pair of polymer strands and using identical orientations as presented in Fig. 1 (g)-(f) of the manuscript, it is evident that eliminating H$_2$O molecules leads to the formation of additional pore channels, designated as 'P', which are otherwise inaccessible in the hydrated form.

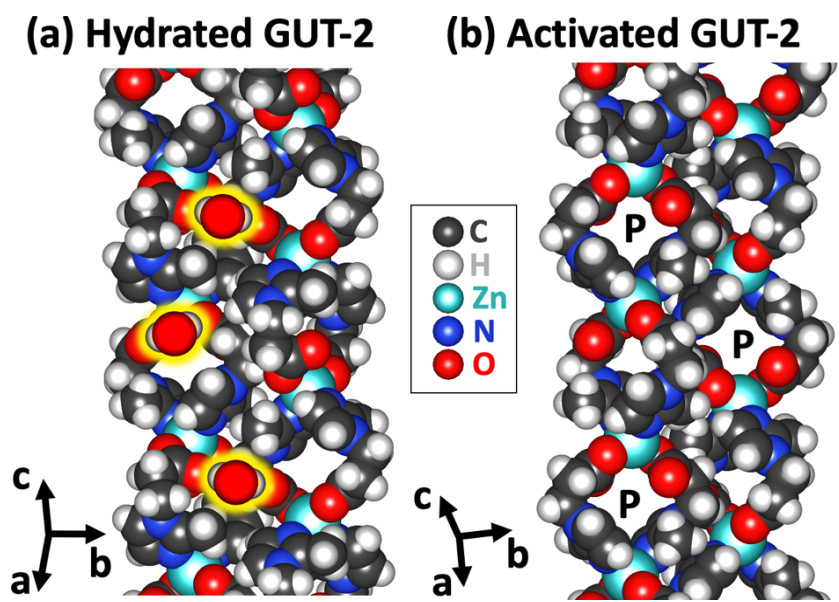

**Fig. S1.** Space-fill models of two polymer strands of hydrated (**a**) and activated GUT-2 (**b**) where two neighboring polymer strands with the perspective oriented along the axis that connects the centers of two neighboring pores. The H$_2$O molecules that are present in the hydrated form have been highlighted in yellow and connect two polymer chains. New pores that become available within the activated structure are labelled with 'P'.

Mercury (version: 2024.2.0)[1] revealed that the hydrated form of GUT-2 contains 1.6 % of the unit cell volume (53.54 Å$^3$) as voids capable of accommodating molecules with a maximum probe radius of 1.2 Å. Larger probes yielded no detectable voids, indicating insufficient space for molecules with a larger probe radius than 1.2 Å. Since H$_2$O is commonly modeled with a probe radius of 1.4 Å in the literature[2], the available voids in hydrated GUT-2 are thus too small to fit any additional H$_2$O molecules. Only the smallest possible molecule, H$_2$, with a probe radius of 1.2 Å, could potentially occupy these voids. A similar situation is observed for the activated form of GUT-2, where 2.0 % of the unit cell volume (32.32 Å$^3$) constitutes of voids of a maximum radius of 1.2 Å.

## S2. Detailed discussion of the single crystal diffraction data for hydrated and activated GUT-2

As discussed in the main text, the crystal structure of activated GUT-2 was successfully solved using single crystal X-ray diffraction, while the structure of hydrated GUT-2, already documented in the literature,[3] was reconfirmed. Therefore, Table S1 provides an in-depth comparison of the crystallographic parameters of both GUT-2 forms.

**Table S1.** Crystal data and structure refinement for the hydrated and activated form of GUT-2 according to the single crystal X-ray diffraction measurements. The parameter Z denotes the number of (chemical) formula units within the unit cell. The value $\rho_{calc}$ represents the calculated crystal density obtained from the unit cell volume and molecular weight. The quantity F(000) represents the total structure factor at zero scattering angle. It corresponds to the sum of the scattering contributions of all atoms in the unit cell and is proportional to the total number of electrons present. The parameter $\mu$ is the linear absorption coefficient, describing the extent to which the crystal absorbs X-rays. The $R_{int}$ value quantifies the internal agreement between symmetry-equivalent reflections and serves as an indicator of data consistency. The refinement residuals, $R_1$ and $wR_2$, provide a measure of the quality of the structural model, comparing observed and calculated diffraction intensities. The cell parameters (*a*, *b* and *c*) following Niggli convention are, thus, provided in brackets.

|  | Hydrated GUT-2 (-173°C)[3] | (1) Hydrated GUT-2 (-173°C) | (2) Activated GUT-2 (100°C) |
|---|---|---|---|
| Formula | $C_{14}H_{18}N_4O_4Zn \cdot H_2O$ | $C_{14}H_{18}N_4O_4Zn \cdot H_2O$ | $C_{14}H_{18}N_4O_4Zn$ |
| Weight [g/mol] | 389.73 | 389.71 | 371.69 |
| Temperature [K] | 100 | 100 | 373 |
| a [Å] | 15.1861(13) | 15.1721(3) | 11.3850(7) [c] |
| b [Å] | 15.0082(13) | 14.9839(3) | 15.2053(7) [b] |
| c [Å] | 15.0568(13) | 15.0445(3) | 9.5687(6) [a] |
| $\alpha = \beta = \gamma$ [°] | 90 | 90 | 90 |
| Volume [Å$^3$] | 3431.7(5) | 3420.17(12) | 1656.46(16) |
| Z | 8 | 8 | 4 |
| $\rho_{calc}$ [g cm$^{-3}$] | 1.509 | 1.514 | 1.490 |
| Crystal system | Orthorhombic | Orthorhombic | Orthorhombic |
| Space group | Pcca | Pcca | Pccn |
| Crystal habit | Block, colourless | Block, colourless | Block, colourless |
| Crystal size [mm$^3$] | 0.05 × 0.05 × 0.04 | 0.17 × 0.12 × 0.09 | 0.17 × 0.12 × 0.09 |
| 2θ range | 2.7–33.1 (Mo K$_\alpha$) | 4.1–76.9 (Cu K$_\alpha$) | 5.8–77.2 (Cu K$_\alpha$) |
| F(000) | 1616 | 1616 | 768 |
| $\mu$ [mm$^{-1}$] | 1.46 | 2.29 | 2.29 |
| $R_{int}$ | 0.063 | 0.034 | 0.10 |
| Independent reflections | 2995 | 3505 | 1735 |
| No. of parameters | 298 | 223 | 107 |
| $R_1$, $wR_2$ (all reflections) | 0.0248, 0.0568 | 0.0426, 0.0815 | 0.1049, 0.2518 |
| $R_1$, $wR_2$ (I $\geq 2\sigma$) | 0.0212, 0.0533 | 0.0344, 0.0785 | 0.0839, 0.2335 |

When evaluating the reliability of a structural model, crystallographers rely on so-called residuals, or R-factors, which quantify the difference between experimental and calculated diffraction data. Overall, lower residuals generally indicate a more reliable and well-refined structural model. These statistical measures help to determine how closely a refined model approximates the actual atomic positions in the crystal. The mathematical foundation of such R-factors is well-established and can be found in textbooks referenced in [4] and [5]. The most common residual, R1, is defined as follows:

$$R_1 = \frac{\sum_{hkl} ||F_{obs}| - |F_{calc}||}{\sum_{hkl} |F_{obs}|}$$

By expressing this value as a percentage, one can easily interpret the agreement between the observed ($F_{obs}$) and calculated ($F_{calc}$) structure factors. However, structural refinement also incorporates weighted contributions, leading to the weighted residual factor, wR$_2$. This parameter is particularly useful in monitoring the impact of adjustments made during the refinement process, as it reflects the minimized quantity in the least-squares refinement. It is defined in the following way:

$$wR_2 = \sqrt{\frac{\sum_{hkl} w(F_{obs}^2 - F_{calc}^2)^2}{\sum_{hkl} w(F_{obs}^2)^2}} \text{ with } w = \frac{1}{\sigma^2 F_{obs}^2}$$

The reason for the index 2 in wR2 originates from the fact that the squared structure factors appear in the equation, thereby giving greater weight to more intense reflections. The weighting factor w ensures that reflections with lower uncertainties contribute more significantly to the refinement process. In some refinement programs, additional terms involving adjustable parameters are introduced to further optimize the weighting scheme.

However, what is perhaps more intriguing is the role of temperature, which, despite not being explicitly present in the equation above, has a profound effect on the refinement results. This enhanced movement makes atomic positions less well-defined, introduces greater uncertainty into the structural refinement and increases mosaicity, which is a measurement of the spread of crystal plane orientations. For the hydrated structure measured at -173°C, the ellipsoids are small and well-contained, indicating minimal atomic motion. In contrast, for the activated structure determined at 100°C, they appear significantly larger, illustrating the pronounced increase in atomic vibrations. Consequently, the observed structure factors deviate more strongly from the calculated ones, leading to an increase in R-values. This is precisely why the structure of activated GUT-2, measured at 100°C, exhibits significantly higher R-values than the low-temperature structure of the hydrated form determined at -173°C.

A visual confirmation of this effect is shown in in Fig. S2 (a)-(b) by comparing thermal ellipsoids – graphical representations of atomic displacement within the crystal lattice – of the determined GUT-2 single crystal structures. Mathematically, thermal ellipsoids are described by the atomic displacement parameters, which define the probability density function for atomic positions. The equation governing their shape is given by:

$$\sum_{i=1}^{3} \sum_{j=1}^{3} U_{ij} x_i x_j = 1$$

Here, $U_{ij}$ represents the elements of the anisotropic displacement tensor and $x_i$ and $x_j$ are atomic displacements along the principal crystallographic axes.

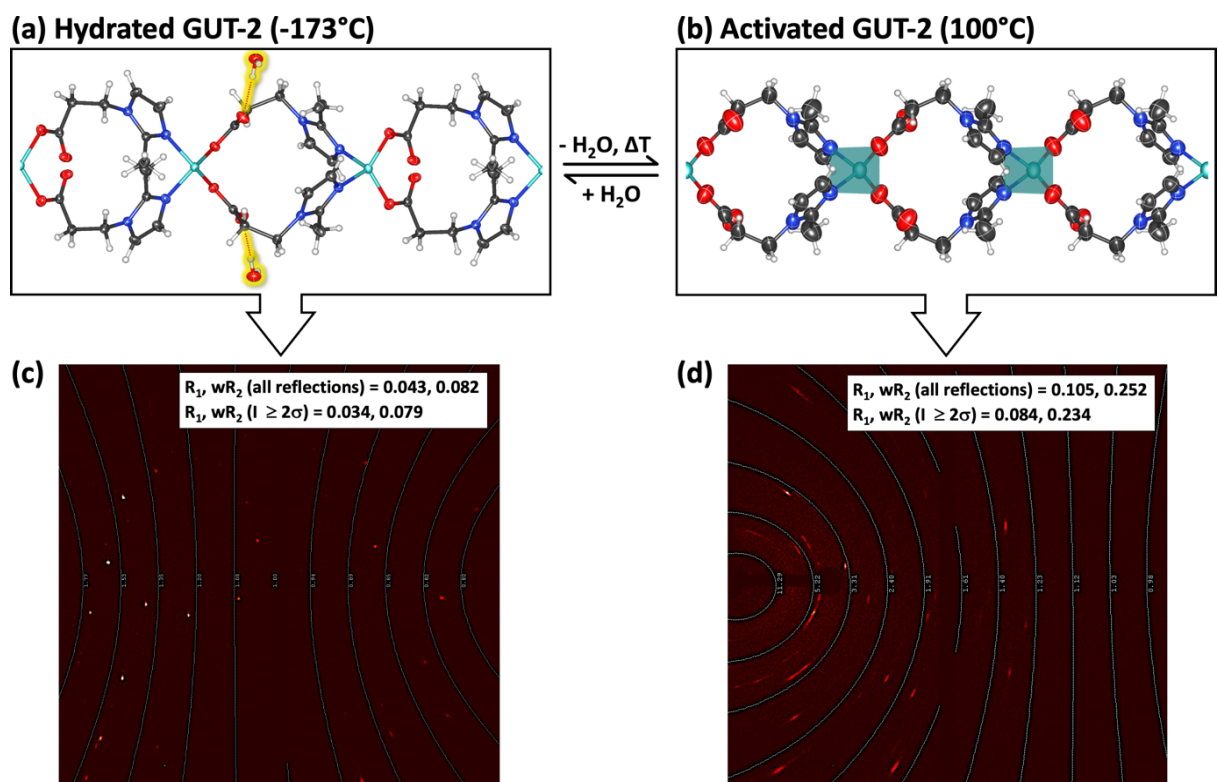

**Fig. S2.** Single crystal structure solution of a single polymer strand of (**a**) hydrated GUT-2 at -173°C and (**b**) activated GUT-2 at 100°C illustrated with thermal ellipsoids representing the displacement parameters of the atoms (projection along a-axis). These thermal ellipsoids are plotted at the 50 % probability level, meaning that each atom is expected to be located within its ellipsoid with this probability. Hydrogen bonding in the hydrated form is indicated using red dashed lines highlighted in yellow that are connected to the H$_2$O molecules also highlighted in yellow. Transparent cyan polyhedrals show the tetrahedral bonding situations around the Zn$^{2+}$ ions. The illustrations visualizing the crystal structures are generated using OLEX2 (version 1.5) [6]. On the bottom the corresponding single crystal X-ray diffraction (XRD) pattern for both GUT-2 states are displayed. While the (**c**) hydrated GUT-2 shows sharp, distinct reflections in its XRD pattern, the (**d**) activated GUT-2 shows more diffuse reflections attributed to the elevated temperature at which the measurement has been performed. To enhance the visibility of these effects, the contrast of the original XRD images has been increased by 50 %.

## S3. Peak position analysis

Tables S2-S5 contain peak positions for key selected diffraction peaks of hydrated and activated GUT-2 for all four temperature measurement series. For the hydrated crystal domains, the (200) diffraction peak offers a reliable measure of coherent crystal size, while the activated domains are characterized through the (110) diffraction peak. The tabulated results illustrate the gradual evolution of peak widths in both hydrated and activated forms of GUT-2.

**Table S2.** Peak positions for the (200) diffraction peak (hydrated GUT-2) and the (110) diffraction peak (activated GUT-2) recorded during the first cooling curve.

|  | Hydrated form, (200) peak | Activated form, (110) peak |
|---|---|---|
| Temperature [°C] | Position [°] | Position [°] |
| 25 | 11.561 | 12.068 |
| 0 | 11.562 | 12.061 |
| -50 | 11.568 | 12.071 |
| -100 | 11.572 | 12.074 |
| -150 | 11.579 | 12.090 |
| -190 | 11.581 | 12.086 |

**Table S3.** Peak positions for the (200) diffraction peak (hydrated GUT-2) and the (110) diffraction peak (activated GUT-2) recorded during the first heating curve.

| | Hydrated form, (200) peak | Activated form, (110) peak |
|---|---|---|
| Temperature [°C] | Position [°] | Position [°] |
| -190 | 11.581 | 12.086 |
| -150 | 11.579 | 12.093 |
| -100 | 11.576 | 12.087 |
| -50 | 11.570 | 12.089 |
| 0 | 11.568 | 12.078 |
| 50 | 11.561 | 12.069 |
| 100 | 11.557 | 12.075 |

**Table S4.** Peak positions for the (110) diffraction peak (activated GUT-2) recorded during the second cooling curve.

| | Activated form, (110) peak |
|---|---|
| Temperature [°C] | Position [°] |
| 100 | 12.086 |
| 50 | 12.094 |
| 0 | 12.097 |
| -50 | 12.100 |
| -100 | 12.107 |
| -150 | 12.117 |
| -180 | 12.126 |

**Table S5.** Peak positions for the (110) diffraction peak (activated GUT-2) recorded during the second heating curve.

| | Activated form, (110) peak |
|---|---|
| Temperature [°C] | Position [°] |
| -180 | 12.126 |
| -150 | 12.117 |
| -100 | 12.111 |
| -50 | 12.104 |
| 0 | 12.103 |
| 50 | 12.094 |
| 100 | 12.088 |
| 150 | 12.081 |
| 200 | 12.069 |
| 250 | 12.057 |

# S4. Instrumental peak broadening

In order to quantify the instrumental broadening effects in PXRD patterns, the standard material LaB$_6$ was measured using the setup from Anton Paar Ltd (as described in Method Section 3.4). Its diffraction pattern is displayed in Fig. S3 (a), with the peak positions full width at half maximum (FWHM) provided in Table S11. Since the X-ray tube uses Cu as the target material, it generates X-rays with two wavelengths, K$_{\alpha 1}$ and K$_{\alpha 2}$, which become distinguishable upon zooming into individual peaks, as shown for the (100) peak in Fig. S3 (b).

Given that LaB$_6$ exhibits its first diffraction peak at approximately 21.4 degrees, and the region of interest for the discussion of the GUT-2 peak widths lies at slightly lower angles, a linear extrapolation was performed on the FWHMs of LaB$_6$ of the K$_{\alpha 1}$ series (see Fig. S3 (c)). In this extrapolation range (from 11.6 degrees to 12.1 degrees), the instrumental FWHM broadening is estimated to be in the range of 0.032 degrees to 0.034 degrees.

**Table S6.** Peak positions and full widths at half maximum (FWHM) for the K$_\alpha$ peak series of LaB$_6$ in the range from 20 degrees to 80 degrees.

| Laue indices | Position [°] | FWHM [°] |
| --- | --- | --- |
| (100) | 21.360 | 0.0351 |
| (110) | 30.388 | 0.0350 |
| (111) | 37.446 | 0.0374 |
| (200) | 43.511 | 0.0370 |
| (210) | 48.962 | 0.0377 |
| (211) | 53.993 | 0.0378 |
| (220) | 63.224 | 0.0414 |
| (300) | 67.551 | 0.0410 |
| (310) | 71.750 | 0.0413 |
| (311) | 75.847 | 0.0440 |

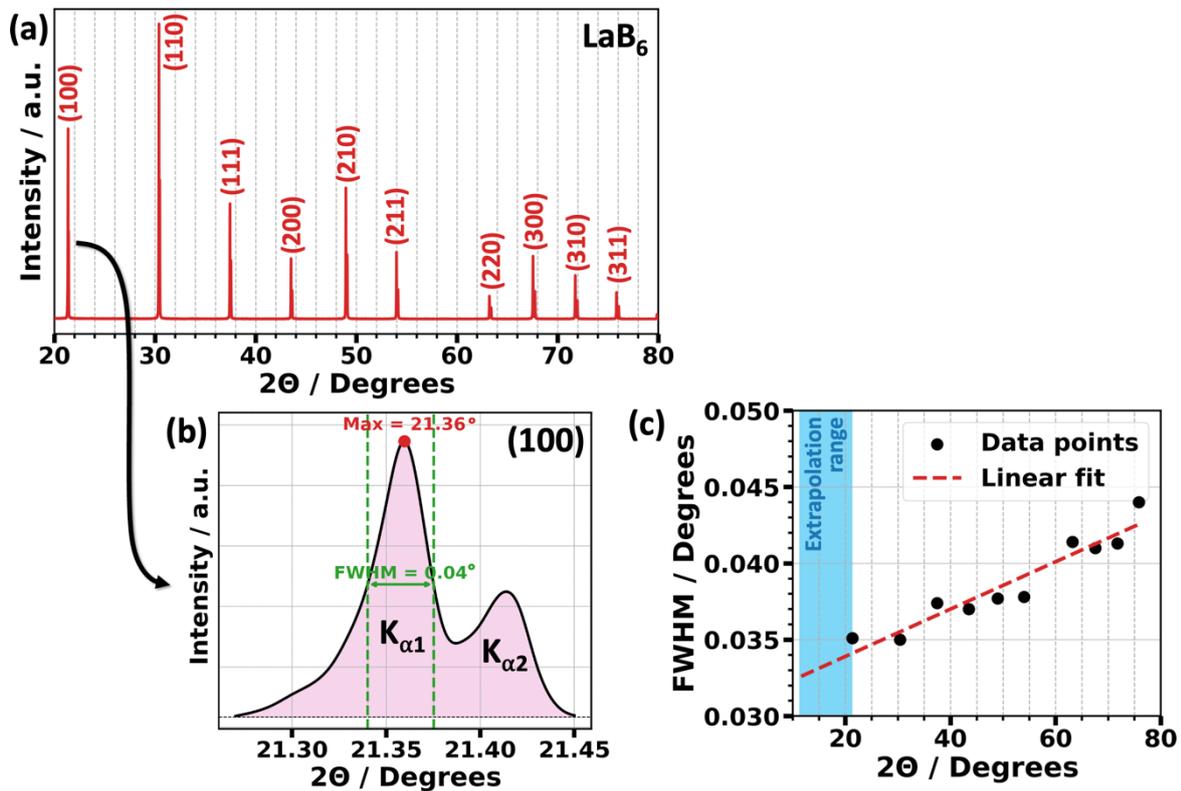

**Fig. S3.** (a) Powder X-ray diffraction pattern of LaB$_6$ from 20 degrees to 80 degrees including the assignment of Laue indices for all peaks. (b) Zoom for the (100) peak of LaB$_6$ and characterization of the K$_{\alpha 1}$ peak with respect to the position of the maximum and the full width at half maximum (FWHM). Linear fits (red dashed line) for the (c) FWHMs (black points) obtained from all K$_{\alpha 1}$ peaks of LaB$_6$ from 20 degrees to 80 degrees. The range of extrapolation is indicated with a blue rectangle, respectively.

## S5. Estimation of crystallite sizes from activated and hydrated GUT-2 powder

From the peak positions reported in Tables S2-S5, the Bragg angles, $\theta$, can be evaluated and from the full width at half maximum (FWHM), denoted as $\beta$, it is possible to estimate the size of the crystallites, $D$, from the Scherrer equation[7] in the following way:

$$D = \frac{K \lambda}{\beta \cos(\theta)}$$

In this equation, the wave length $\lambda$ is 1.5418 Å according to the instrument settings and $K$, which is the shape factor, is assumed to be 0.89. The value for $K$ stems from the original derivation of the Scherrer equation[7] that assumed crystallites with cubic symmetry. Given that GUT-2 possesses an orthorhombic crystal system with unit cell parameters close to cubic symmetry, this $K$ value provides a reasonable assumption for the crystallite size calculations.

The FWHM values are calculated as the peak width at an intensity that corresponds to fifty percent of its highest value Since XRD measurements record a combined signal of both K$\alpha_1$ and K$\alpha_2$ radiation, the overlap of these lines can lead to broadened peaks, affecting the accuracy of the crystal size analysis. Because of that, the Rachinger correction[8] (which is available in the program X'Pert HighScore Plus[9]) was applied before determining the FWHM values to remove the K$\alpha_2$ lines. The estimated crystallite sizes for the hydrated and activated form of GUT-2 are reported in Tables S7-S10. A detailed discussion of these values can be found in the paper.

**Table S7.** Full width at half maximum (FWHM) and estimated crystallite sizes using the Scherrer equation[7] for the (200) diffraction peak (hydrated GUT-2) and the (110) diffraction peak (activated GUT-2) recorded during the first cooling curve.

| Temperature [°C] | Hydrated form, (200) peak | | Activated form, (110) peak | |
| --- | --- | --- | --- | --- |
| | FWHM [°] | Crystallite size [nm] | FWHM [°] | Crystallite size [nm] |
| 25 | 0.0684 | 116 | 0.1278 | 62 |
| 0 | 0.0742 | 107 | 0.1288 | 62 |
| -50 | 0.0692 | 114 | 0.1248 | 63 |
| -100 | 0.0736 | 108 | 0.1237 | 64 |
| -150 | 0.0736 | 108 | 0.1237 | 64 |
| -190 | 0.0702 | 113 | 0.1239 | 64 |
| | Average size [nm] = 111 ± 4 | | Average size [nm] = 63 ± 1 | |

**Table S8.** Full width at half maximum (FWHM) and estimated crystallite sizes using the Scherrer equation[7] for the (200) diffraction peak (hydrated GUT-2) and the (110) diffraction peak (activated GUT-2) recorded during the first heating curve.

| Temperature [°C] | Hydrated form, (200) peak | | Activated form, (110) peak | |
| --- | --- | --- | --- | --- |
| | FWHM [°] | Crystallite size [nm] | FWHM [°] | Crystallite size [nm] |
| -190 | 0.0702 | 113 | 0.1239 | 64 |
| -150 | 0.0735 | 108 | 0.1019 | 78 |
| -100 | 0.0698 | 113 | 0.1191 | 67 |
| -50 | 0.0687 | 115 | 0.1190 | 67 |
| 0 | 0.0705 | 112 | 0.1209 | 66 |
| 50 | 0.0717 | 111 | 0.1151 | 69 |
| 100 | 0.0754 | 105 | 0.1084 | 73 |
| | Average size [nm] = 111 ± 3 | | Average size [nm] = 69 ± 5 | |

**Table S9.** Full width at half maximum (FWHM) and estimated crystallite sizes using the Scherrer equation[7] for the (110) diffraction peak (activated GUT-2) recorded during the second cooling curve.

| Temperature [°C] | Activated form, (110) peak | |
| --- | --- | --- |
| | FWHM [°] | Crystallite size [nm] |
| 100 | 0.1084 | 73 |
| 50 | 0.1016 | 78 |
| 0 | 0.1035 | 76 |
| -50 | 0.1039 | 76 |
| -100 | 0.1045 | 76 |
| -150 | 0.1050 | 75 |
| -180 | 0.1046 | 76 |
| | Average size [nm] = 76 ± 2 | |

**Table S10.** Full width at half maximum (FWHM) and estimated crystallite sizes using the Scherrer equation[7] for the (110) diffraction peak (activated GUT-2) recorded during the second heating curve.

| Temperature [°C] | Activated form, (110) peak | |
| --- | --- | --- |
| | FWHM [°] | Crystallite size [nm] |
| -180 | 0.1046 | 76 |
| -150 | 0.1050 | 75 |
| -100 | 0.1041 | 76 |
| -50 | 0.1015 | 78 |
| 0 | 0.1040 | 76 |
| 50 | 0.1040 | 76 |
| 100 | 0.1042 | 76 |
| 150 | 0.1022 | 78 |
| 200 | 0.1030 | 77 |
| 250 | 0.1019 | 78 |
| | Average size [nm] = 77 ± 1 | |

## S6. Temperature-dependent cell parameters of hydrated and activated GUT-2 obtained from Rietveld refinements

To determine temperature-dependent cell parameters and derive direction-specific thermal expansion coefficients, Rietveld refinements were performed. The outcomes of these Rietveld refinements for the four sections of the temperature cycle shown in Fig. 2 (a) are detailed in Fig. S4–S9. The data reveal that, for the hydrated form of GUT-2 in C1 and H1, not all peaks are accurately captured using only the hydrated structure model. In fact, it turned out that incorporating also the activated structure model of GUT-2 in the refinement process is necessary to account for all experimental peaks (for details see main manuscript).

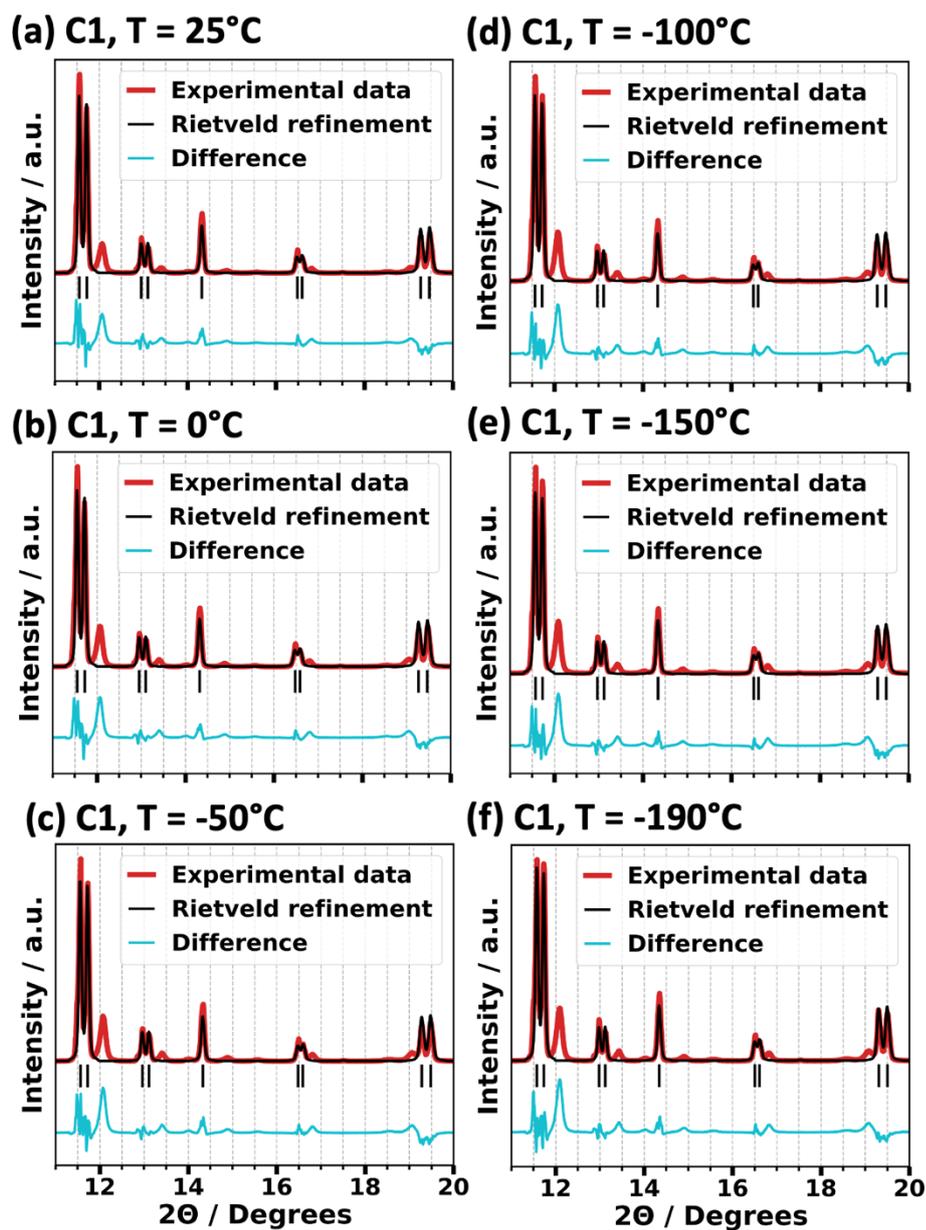

**Fig. S4.** Overlay of the temperature-dependent powder X-ray diffraction pattern (red curves) and the Rietveld refinements (black curves) based on the hydrated form of GUT-2 for the first cooling cycle (C1). The calculated Bragg peaks are shown as vertical lines. The difference between the experimental data and the Rietveld refinement is shown as a blue curve below in each of the subplots.

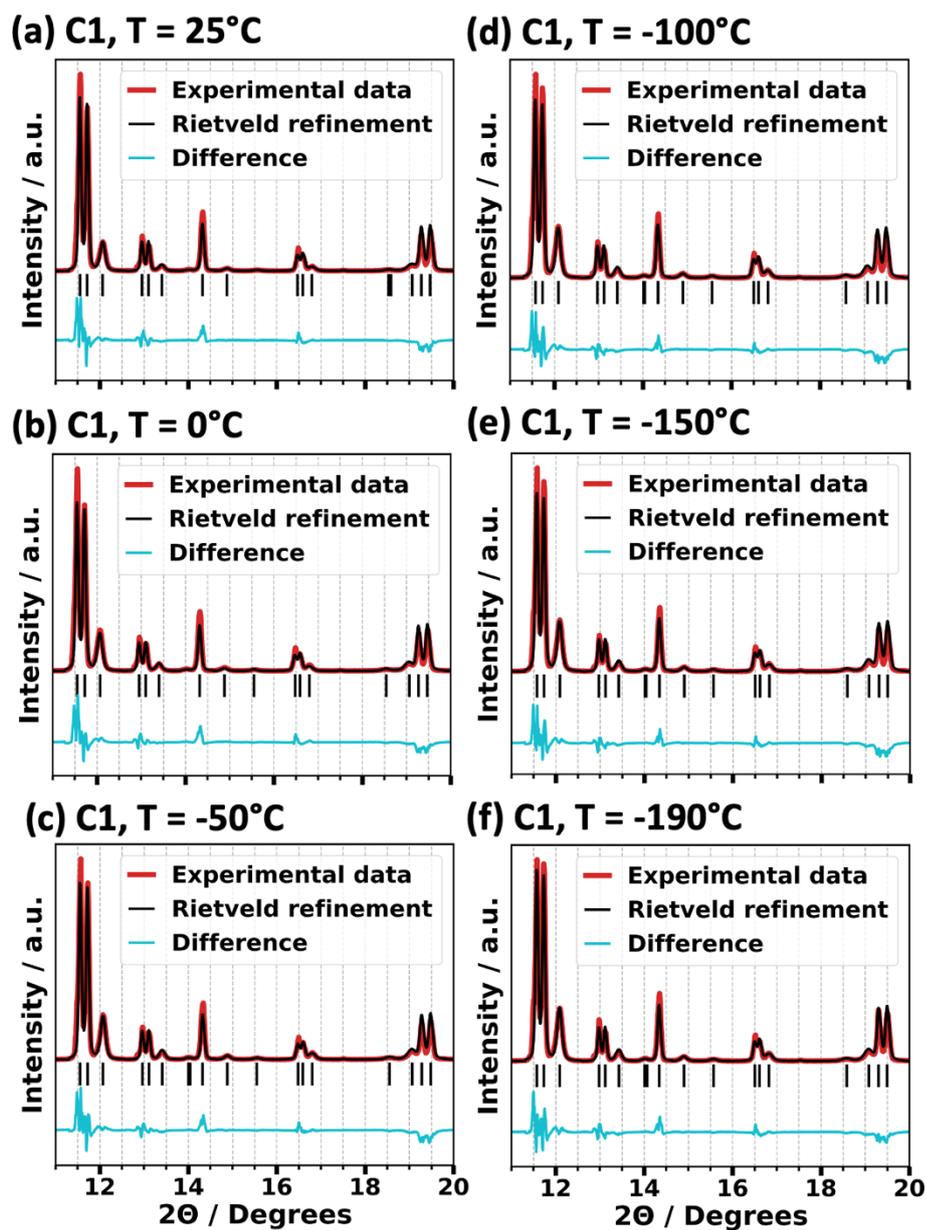

**Fig. S5.** Overlay of the temperature-dependent powder X-ray diffraction pattern (red curves) and the Rietveld refinements (black curves) based on the hydrated and activated form of GUT-2 for the first cooling cycle (C1). The calculated Bragg peaks are shown as vertical lines. The difference between the experimental data and the Rietveld refinement is shown as a blue curve below in each of the subplots.

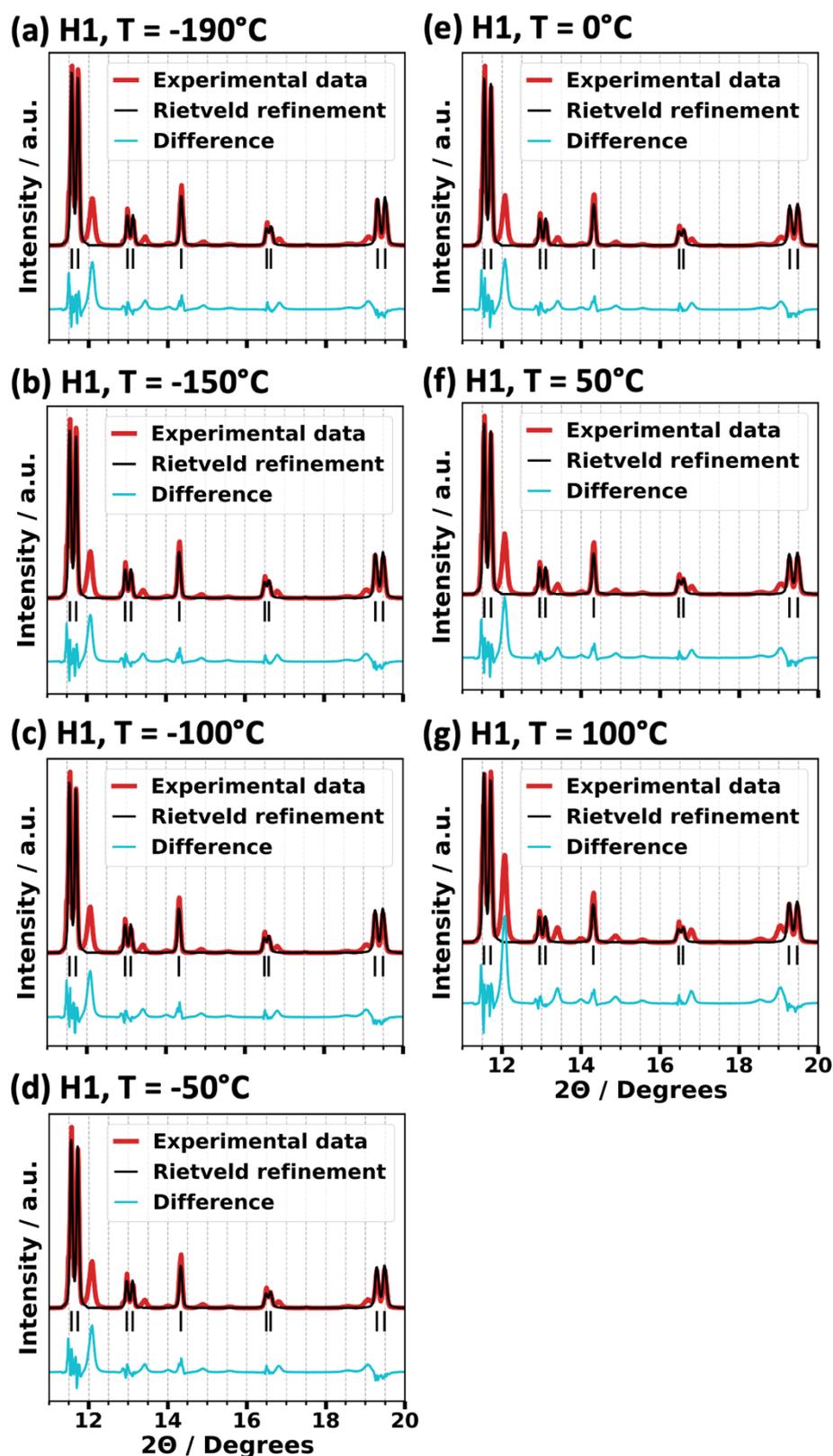

**Fig. S6.** Overlay of the temperature-dependent powder X-ray diffraction pattern (red curves) and the Rietveld refinements (black curves) based on the hydrated form of GUT-2 for the first heating cycle (H1). The calculated Bragg peaks are shown as vertical lines. The difference between the experimental data and the Rietveld refinement is shown as a blue curve below in each of the subplots.

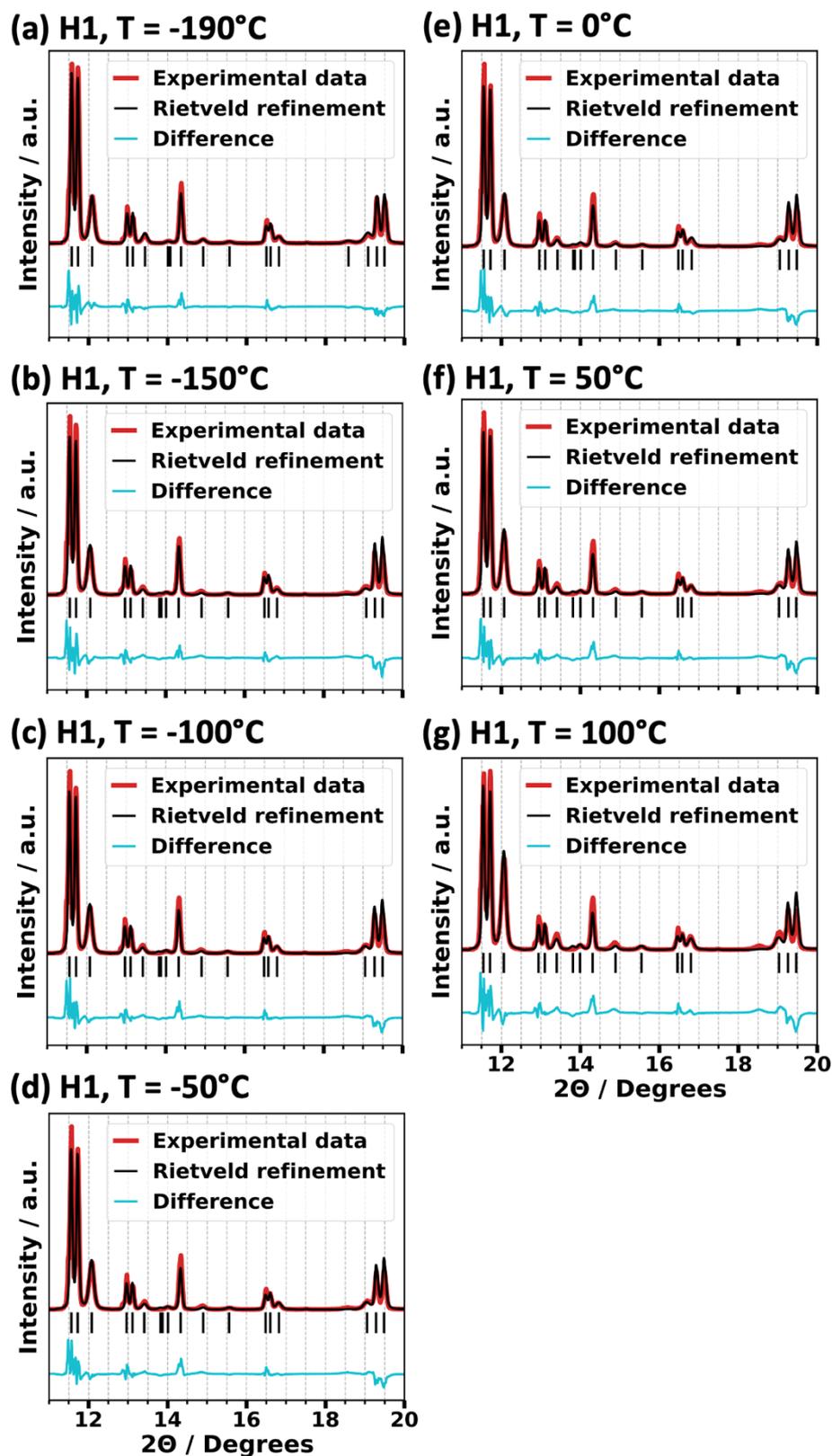

**Fig. S7.** Overlay of the temperature-dependent powder X-ray diffraction pattern (red curves) and the Rietveld refinements (black curves) based on the hydrated and activated form of GUT-2 for the first heating cycle (H1). The calculated Bragg peaks are shown as vertical lines. The difference between the experimental data and the Rietveld refinement is shown as a blue curve below in each of the subplots.

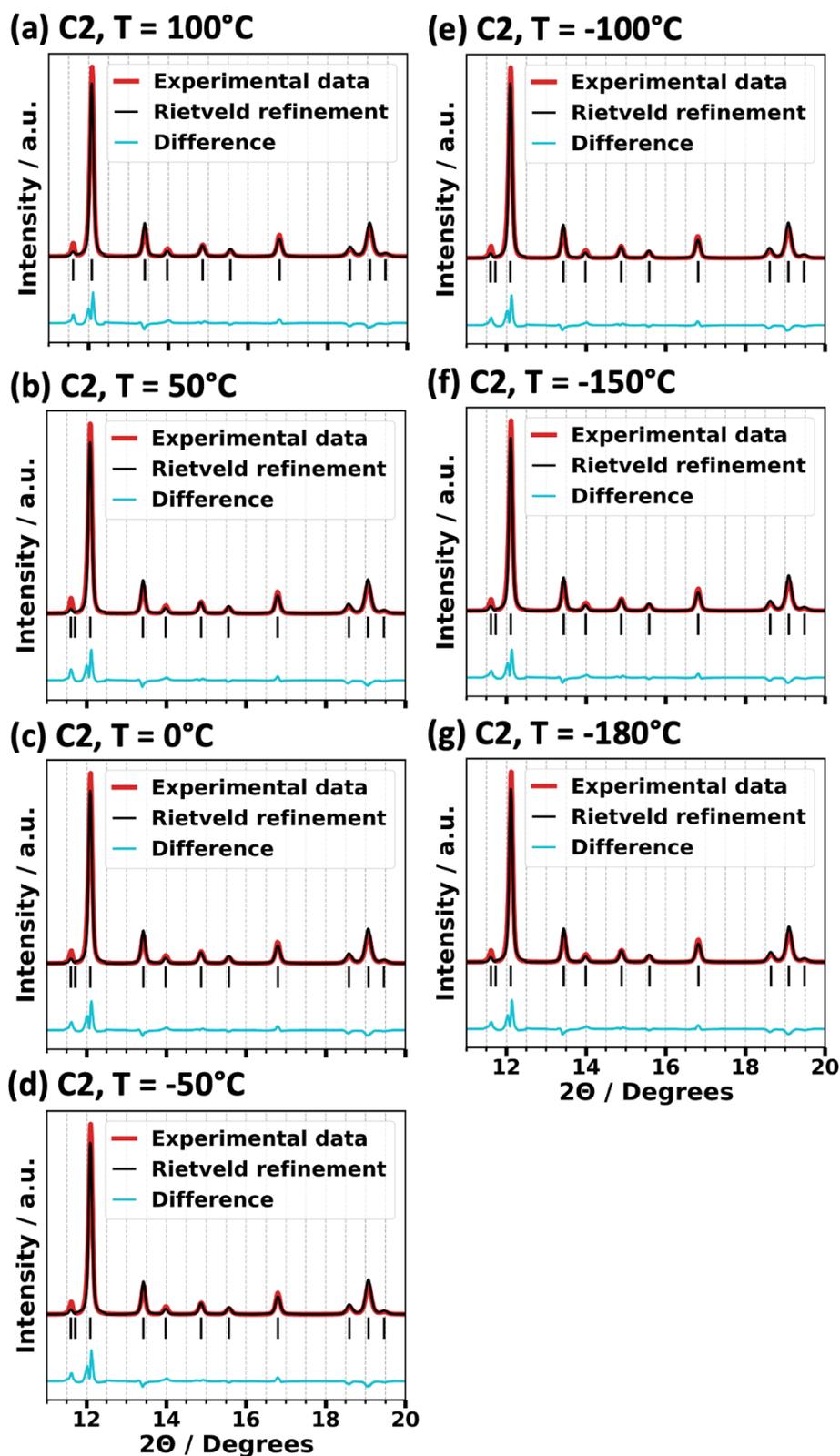

**Fig. S8.** Overlay of the temperature-dependent powder X-ray diffraction pattern (red curves) and the Rietveld refinements (black curves) based on the activated form of GUT-2 for the second cooling cycle (C2). The calculated Bragg peaks are shown as vertical lines. The difference between the experimental data and the Rietveld refinement is shown as a blue curve below in each of the subplots.

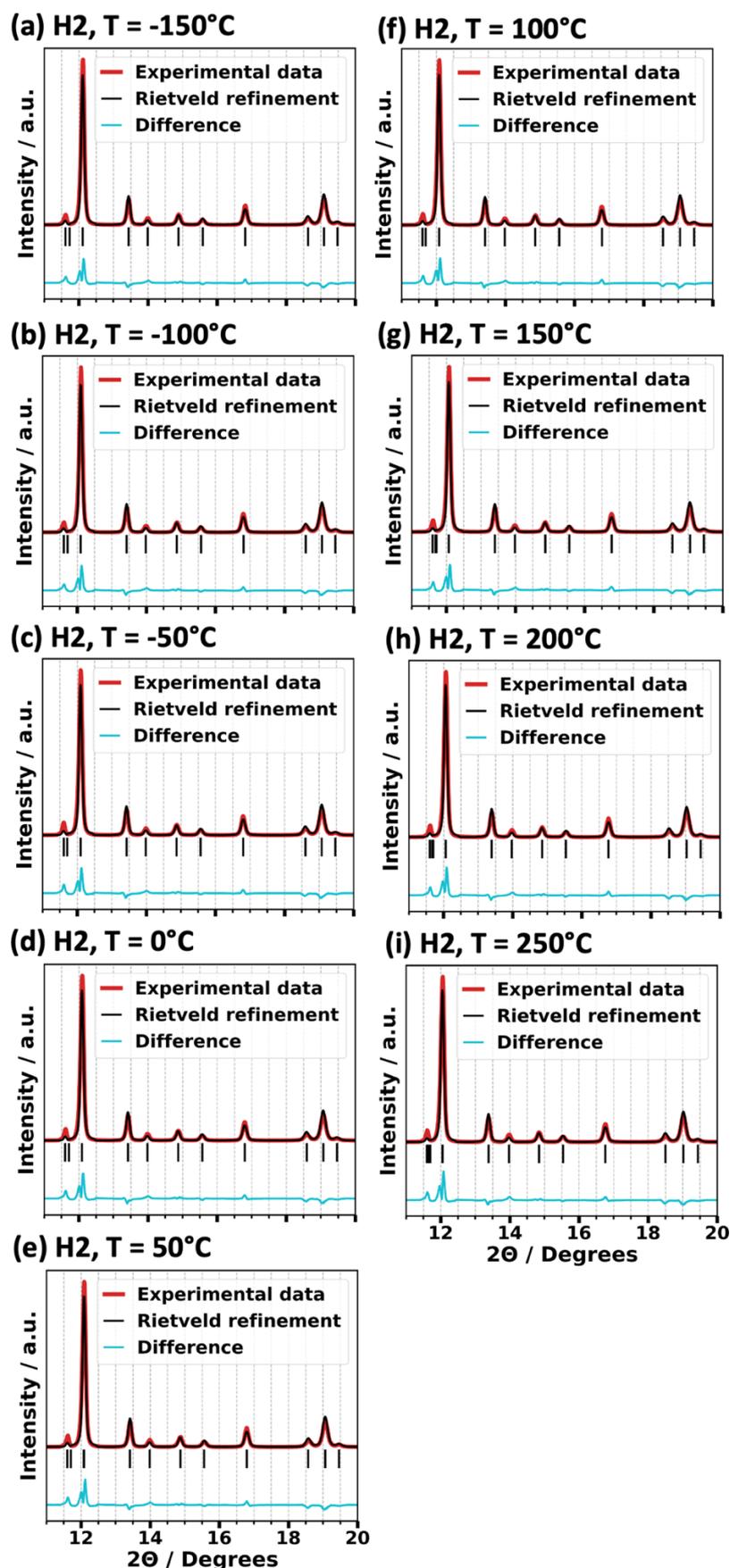

**Fig. S9.** Overlay of the temperature-dependent powder X-ray diffraction pattern (red curves) and the Rietveld refinements (black curves) based on the activated form of GUT-2 for the second heating cycle (H2). The calculated Bragg peaks are shown as vertical lines. The difference between the experimental data and the Rietveld refinement is shown as a blue curve below in each of the subplots.

Three statistical parameters, $R_p$, $R_{wp}$, and $R_{exp}$ are commonly used in Rietveld refinement to measure how well the observed diffraction pattern matches the calculated pattern from the model, allowing for an evaluation of the quality of the refinement. $R_p$ is called the R-pattern or profile residual and compares the difference between the observed intensity $y_i$ and the calculated intensity $y_c$ for each data point in the diffraction pattern. The closer $R_p$ is to zero, the better the fit.

$$R_p = \frac{\sum |y_i - y_c|}{\sum y_i}$$

$R_{wp}$ is a weighted version of $R_p$ and is, therefore, called the weighted profile residual. It takes into account the uncertainty in the observed intensities ($I_{obs}$), typically giving more weight $w_i$ to data points with higher intensities (lower noise). Here, $\sigma$ is the standard deviation of the observed intensity, which is typically determined from counting statistics. In diffraction experiments, the uncertainty in measured (observed) intensities follows Poisson statistics.

$$R_{wp} = \sqrt{\frac{\sum w_i(y_i - y_c)^2}{\sum w_i y_i^2}} \text{ with } w_i = \frac{1}{\sigma_i^2} \text{ where } \sigma(I_{obs}) = \sqrt{I_{obs}}$$

Finally, the expected R-factor, $R_{exp}$, provides a measure of how good the fit could be based purely on the level of noise. It depends on the number of data points $N$ and on the number of parameters that are refined denoted with $P$.

$$R_{exp} = \sqrt{\frac{(N-P)}{\sum w_i y_i^2}}$$

A summary of the three statistical parameters $R_p$, $R_{wp}$ and $R_{exp}$ alongside with the temperature-dependent cell parameters $a$, $b$ and $c$ together with the volume $V$ can be found in Tables S11-S14 for all measured points on both cooling and heating curves.

**Table S11.** Temperature-dependent cell parameters of hydrated GUT-2 for the first cooling curve. $R_p$ (R-pattern), $R_{wp}$ (R-weighted pattern) and $R_{exp}$ (R-expected) are the standard reliability factors used in Rietveld refinement to assess the quality of the fit between the observed and calculated diffraction patterns. $R_p^*$ is obtained using both the activated and hydrated form simultaneously for the Rietveld refinement.

| T [°C] | a [Å] | b [Å] | c [Å] | V [Å³] | $R_p^*$ [%] | $R_p$ [%] | $R_{wp}$ [%] | $R_{exp}$ [%] |
|---|---|---|---|---|---|---|---|---|
| 25 | 15.2901 | 15.0972 | 15.0728 | 3479.36 | 18.85 | 28.03 | 38.33 | 2.23 |
| 0 | 15.2895 | 15.0980 | 15.0725 | 3479.34 | 18.36 | 29.96 | 41.76 | 2.21 |
| -50 | 15.2874 | 15.0985 | 15.0713 | 3478.71 | 18.46 | 31.18 | 43.79 | 2.19 |
| -100 | 15.2792 | 15.0920 | 15.0676 | 3474.52 | 18.17 | 31.58 | 44.43 | 2.19 |
| -150 | 15.2711 | 15.0848 | 15.0624 | 3469.80 | 18.17 | 36.09 | 51.49 | 2.20 |
| -190 | 15.2601 | 15.0790 | 15.0560 | 3464.36 | 16.33 | 30.33 | 43.76 | 2.20 |

**Table S12.** Temperature-dependent cell parameters of hydrated GUT-2 for the first heating curve. $R_p$ (R-pattern), $R_{wp}$ (R-weighted pattern) and $R_{exp}$ (R-expected) are the standard reliability factors used in Rietveld refinement to assess the quality of the fit between the observed and calculated diffraction

patterns. $R_p^*$ is obtained using both the activated and hydrated form simultaneously for the Rietveld refinement.

| T [°C] | a [Å] | b [Å] | c [Å] | V [Å³] | $R_p^*$ [%] | $R_p$ [%] | $R_{wp}$ [%] | $R_{exp}$ [%] |
|---|---|---|---|---|---|---|---|---|
| -190 | 15.2601 | 15.0790 | 15.0560 | 3464.36 | 16.33 | 30.33 | 43.76 | 2.20 |
| -150 | 15.2638 | 15.0816 | 15.0574 | 3466.25 | 18.81 | 30.46 | 43.81 | 2.20 |
| -100 | 15.2718 | 15.0895 | 15.0628 | 3471.13 | 18.82 | 30.30 | 43.73 | 2.18 |
| -50 | 15.2766 | 15.0933 | 15.0645 | 3473.49 | 19.01 | 30.64 | 44.05 | 2.18 |
| 0 | 15.2826 | 15.0969 | 15.0665 | 3476.13 | 20.14 | 31.64 | 45.02 | 2.18 |
| 50 | 15.2903 | 15.1036 | 15.0701 | 3480.28 | 20.02 | 33.95 | 47.21 | 2.18 |
| 100 | 15.2947 | 15.1093 | 15.0721 | 3483.05 | 21.45 | 39.39 | 52.96 | 2.18 |

**Table S13.** Temperature-dependent cell parameters of activated GUT-2 for the second cooling curve $R_p$ (R-pattern), $R_{wp}$ (R-weighted pattern) and $R_{exp}$ (R-expected) are the standard reliability factors used in Rietveld refinement to assess the quality of the fit between the observed and calculated diffraction patterns.

| T [°C] | a [Å] | b [Å] | c [Å] | V [Å³] | $R_p$ [%] | $R_{wp}$ [%] | $R_{exp}$ [%] |
|---|---|---|---|---|---|---|---|
| 100 | 11.3682 | 15.2203 | 9.5433 | 1651.17 | 17.62 | 21.76 | 2.16 |
| 50 | 11.3686 | 15.2186 | 9.5407 | 1649.67 | 17.38 | 22.96 | 2.16 |
| 0 | 11.3654 | 15.2181 | 9.5357 | 1649.29 | 17.90 | 22.15 | 2.16 |
| -50 | 11.3631 | 15.2174 | 9.5311 | 1648.09 | 17.70 | 22.00 | 2.16 |
| -100 | 11.3596 | 15.2166 | 9.5247 | 1646.37 | 17.48 | 21.77 | 2.16 |
| -150 | 11.3538 | 15.2126 | 9.5145 | 1643.35 | 17.44 | 21.67 | 2.17 |
| -180 | 11.3503 | 15.2118 | 9.5087 | 1641.77 | 17.22 | 21.37 | 2.17 |

**Table S14.** Temperature-dependent cell parameters of activated GUT-2 for the second heating curve. $R_p$ (R-pattern), $R_{wp}$ (R-weighted pattern) and $R_{exp}$ (R-expected) are the standard reliability factors used in Rietveld refinement to assess the quality of the fit between the observed and calculated diffraction patterns.

| T [°C] | a [Å] | b [Å] | c [Å] | V [Å³] | $R_p$ [%] | $R_{wp}$ [%] | $R_{exp}$ [%] |
|---|---|---|---|---|---|---|---|
| -180 | 11.3503 | 15.2118 | 9.5087 | 1641.77 | 17.22 | 21.37 | 2.17 |
| -150 | 11.3548 | 15.2156 | 9.5150 | 1643.91 | 17.24 | 21.39 | 2.17 |
| -100 | 11.3594 | 15.2185 | 9.5232 | 1646.31 | 17.40 | 21.56 | 2.17 |
| -50 | 11.3624 | 15.2186 | 9.5288 | 1647.71 | 17.64 | 22.16 | 2.16 |
| 0 | 11.3658 | 15.2202 | 9.5363 | 1649.68 | 17.93 | 22.50 | 2.16 |
| 50 | 11.3684 | 15.2198 | 9.5418 | 1650.96 | 17.60 | 22.13 | 2.16 |
| 100 | 11.3726 | 15.2217 | 9.5486 | 1652.98 | 17.82 | 22.37 | 2.16 |
| 150 | 11.3764 | 15.2228 | 9.5585 | 1655.35 | 17.90 | 22.53 | 2.16 |
| 200 | 11.3840 | 15.2240 | 9.5672 | 1658.09 | 18.22 | 22.86 | 2.16 |
| 250 | 11.3912 | 15.2261 | 9.5803 | 1661.63 | 18.35 | 23.14 | 2.15 |

## S7. Test runs causing partially activation of hydrated GUT-2

Before recording the powder X-ray diffraction (PXRD) pattern of the actual cooling curve C1, reported in the main manuscript, two preliminary test runs were conducted for the first cooling cycle, whose temperature-dependent PXRD pattern are given in Fig. S10 (a)-(b). Phase quantification through Rietveld refinements shown in Fig. S11 (a)-(b) indicates a partial activation of the powder sample during these preliminary tests. Specifically, each time the sample was cooled from 25°C to 0°C, approximately 10 % activation was observed, which we attribute to the fact that the experiments were performed in vacuum. Consequently, in the temperature-dependent PXRD pattern, features of the activated GUT-2 form are present. This is seen, for example, for the at 12.1° and indicates an activation level of around 30 %. As shown in Fig. S11 (a), phase quantification of the first heating curve reveals that the sample starts at 30 % activation. Significant further activation of the GUT-2 powder only begins after approximately two hours, when the temperature reaches around 50°C.

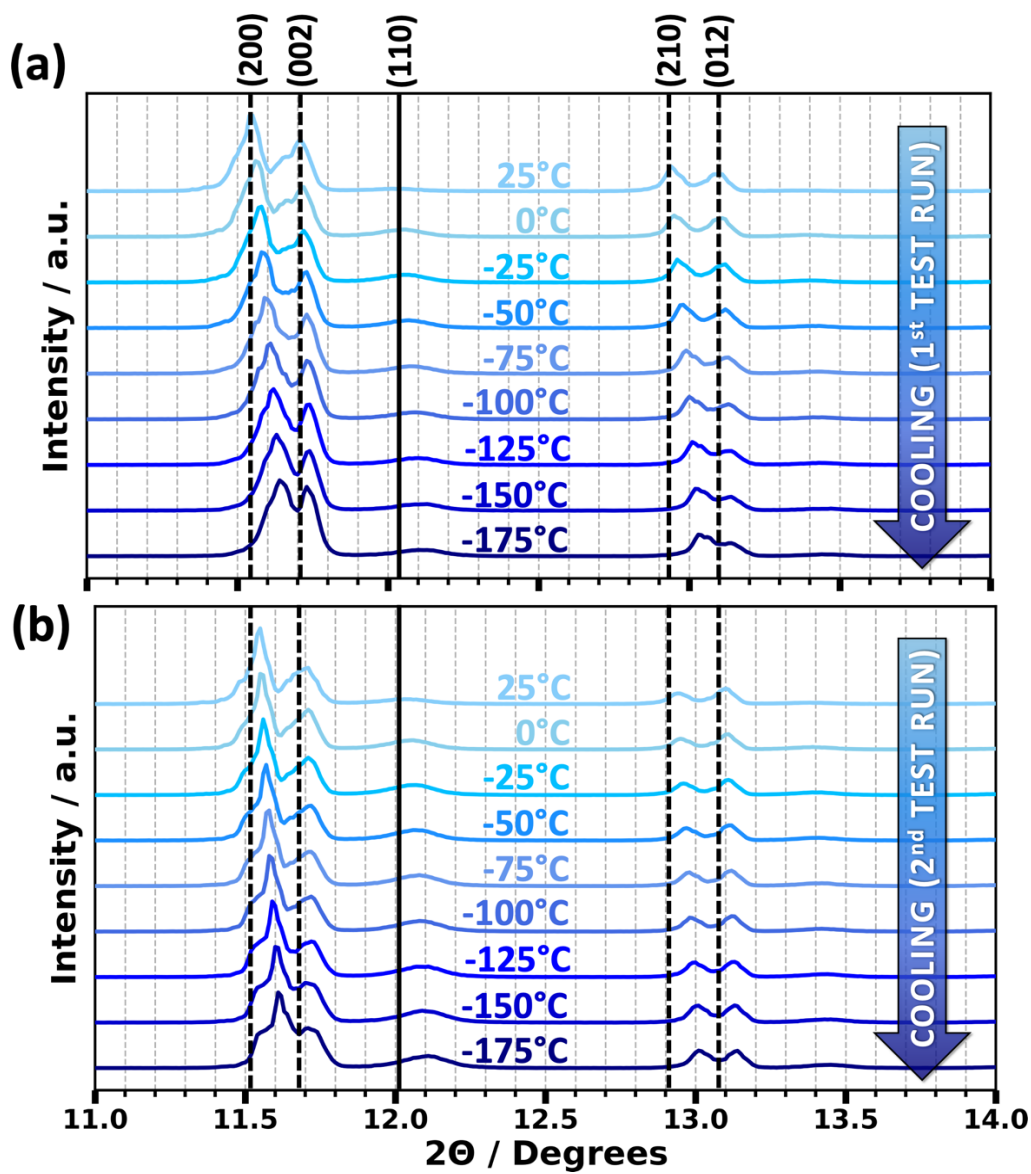

**Fig. S10.** (**a**)-(**b**) Temperature-dependent powder X-ray diffraction pattern in the range from 11 to 14 degrees for two cooling cycle test runs of GUT-2. Peaks associated with hydrated GUT-2 are highlighted by dashed vertical lines, whereas solid vertical lines indicate the peaks of the activated form.

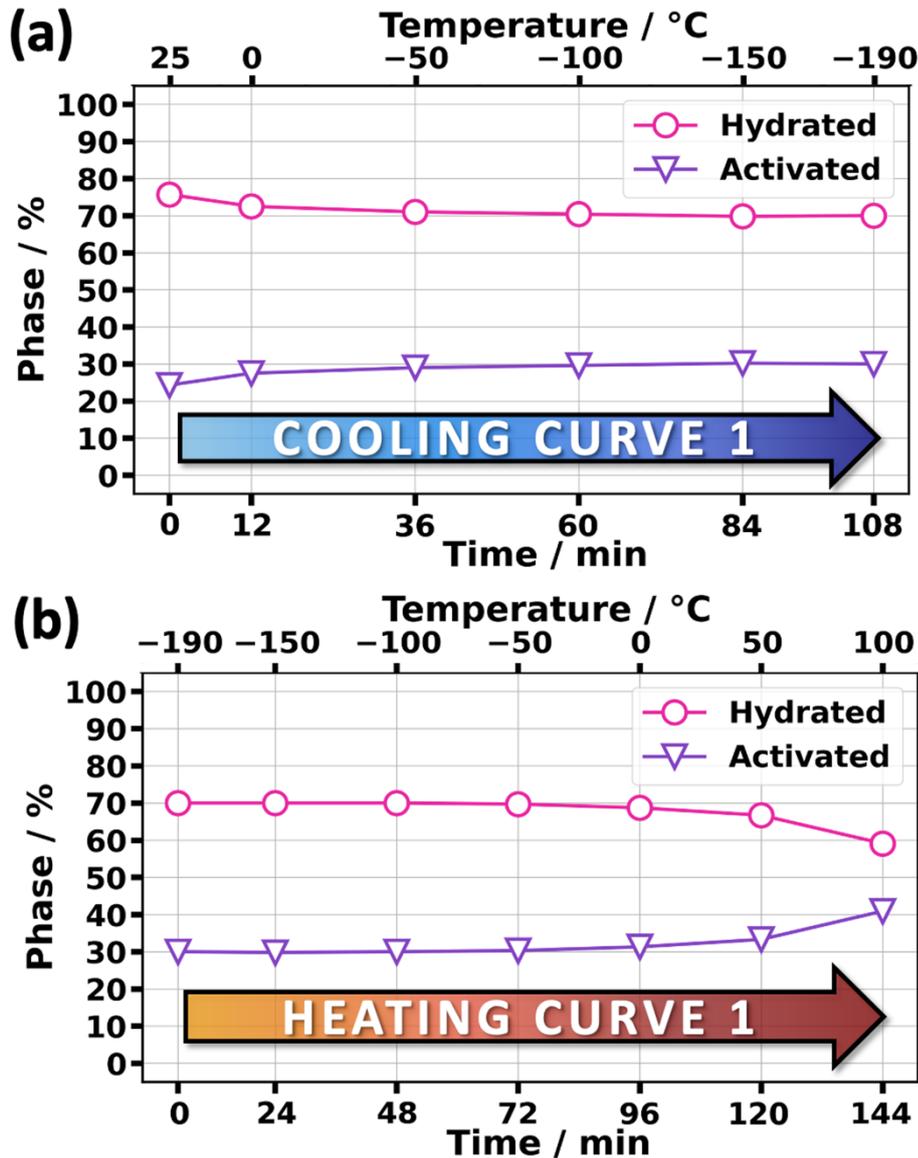

**Fig. S11.** Phase quantifications for the first cooling (**a**) and first heating curve (**b**) with respect to temperature based on Rietveld refinements. The data points for the amount of hydrated form are shown as pink circles (O), while the data points for the percentage of activated phase are represented as purple, inverted triangles (∇).

### S8. Thermal expansion and bonding strengths

The thermal expansion of materials is fundamentally linked to the type of interatomic interactions. A useful way to describe these interactions is to perform the Taylor expansion of a general anharmonic potential around its minimum (as shown for example in the solid state physics textbooks by Gross and Marx[10] or by Kittel[11]), given by:

$$U(x) = cx^2 - gx^3$$

where $x$ represents the atomic displacement and the coefficients $c$ and $g$ determine the curvature and the asymmetry of the potential well. This polynomial approximation captures the essential anharmonic effects. Unlike the Morse potential, which has an exponential decay for large bond lengths, this polynomial expansion does not describe any bond breaking.

To derive the thermal expansion from the potential energy function $U(x)$, we can use the Boltzmann probability distribution to calculate the expectation value of the atomic displacement, denoted as $\langle x \rangle$ with $\beta = 1/k_BT$ ($k_B$ refers to as the Boltzmann constant, $T$ is the temperature) and expanding $e^{-\beta f(x)}$ using a Taylor series around the harmonic term[11]:

$$\langle x \rangle = \frac{\int_{-\infty}^{\infty} x e^{-\beta U(x)} dx}{\int_{-\infty}^{\infty} e^{-\beta U(x)} dx} \cong \frac{\int_{-\infty}^{\infty} e^{-\beta c x^2}(x + \beta g x^4) dx}{\int_{-\infty}^{\infty} e^{-\beta c x^2} dx} = \frac{\frac{3\pi^{1/2}}{4} \frac{g}{c^{5/2}} \beta^{-3/2}}{\left(\frac{\pi}{\beta c}\right)^{1/2}} = \frac{3g}{4c^2} k_B T$$

From this result, we see that the cubic and quartic terms of U(x) introduce thermal expansion. This means that for a harmonic oscillator ($g = 0$), $\langle x \rangle$ remains constant and thus this model is unable to describe any thermal expansion. In contrast, as soon as anharmonic terms ($g \neq 0$) are present, $\langle x \rangle$ changes with temperature, leading to thermal expansion. Moreover, the coefficient $c$ defines the curvature at the minimum of the potential well and is directly linked to the bond stiffness. The stronger the chemical bond, the higher $c$ becomes the smaller thermal movement of the atoms. Therefore, as long as $g$ does not increase at least with $c$-squared, a more strongly bonding potential is expected to result in a reduced thermal expansion.

## S9. Kinetics of H₂O adsorption/desorption

In the course of experiments, we investigated the kinetics of $H_2O$ adsorption and desorption of GUT-2 in greater detail. The temperature-dependent PXRD patterns and corresponding phase quantifications of these kinetic studies are displayed in Fig. S12. As outlined in the manuscript, hydrated GUT-2 undergoes activation relatively easily: When the sample is heated to 90°C, full activation occurs within approximately 30 minutes, as evidenced by a pronounced alteration in the PXRD pattern. In a second experiment, heating was limited to 50°C, leading to full activation after roughly 6 hours.

We also examined the rehydration process (Fig. S12 (c)) by exposing activated GUT-2 to standard laboratory conditions (30 % relative humidity). Here, $H_2O$ adsorption occurs much more slowly, beginning after several hours with a decrease in the 12.1 degrees peak intensity in the PXRD pattern. Complete rehydration, however, requires at least two days (see Fig. S12 (f)), though the process of $H_2O$ ad- and desorption remains fully reversible keeping the framework intact.

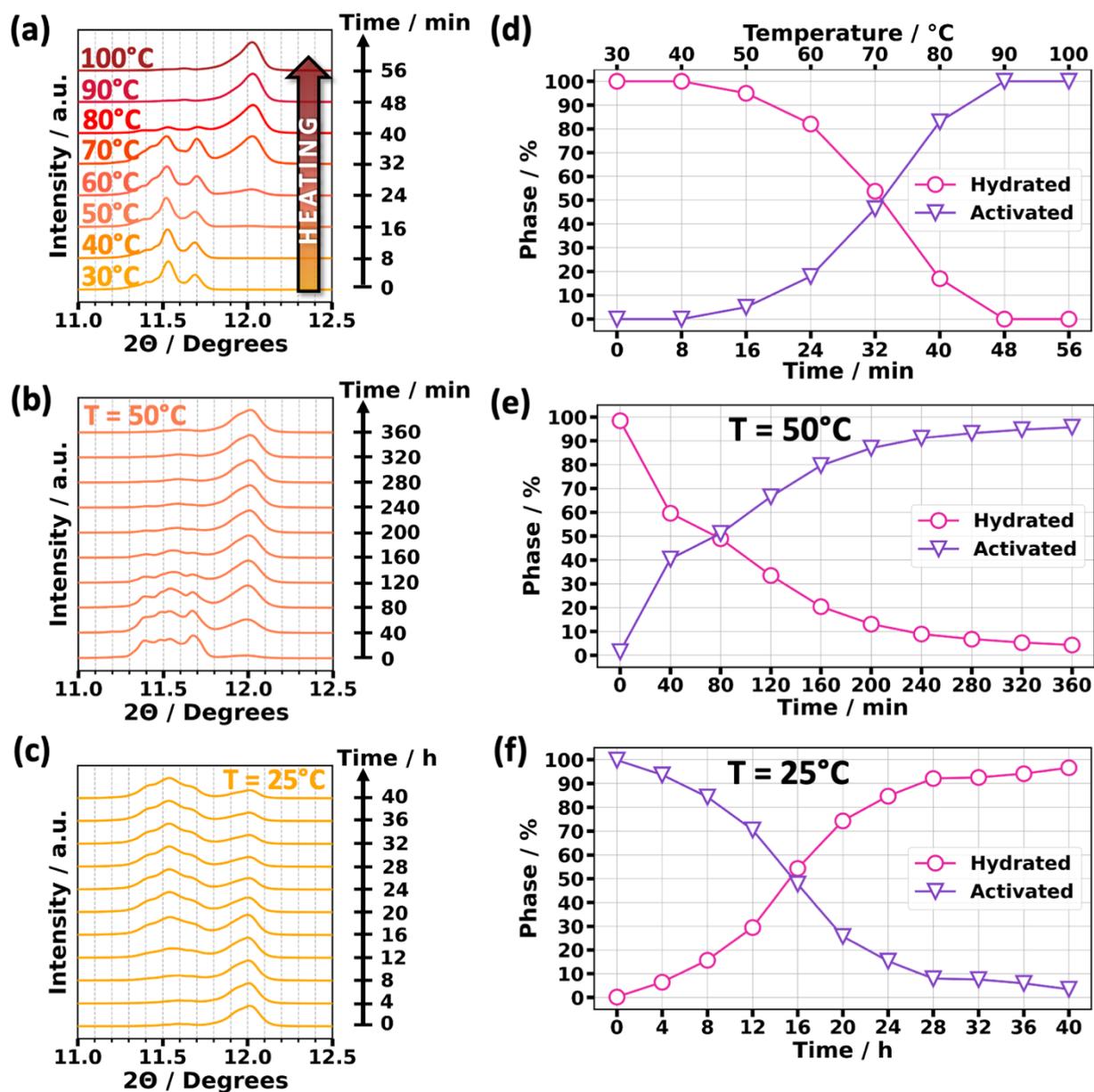

**Fig. S12.** Experimental powder X-ray diffraction pattern (PXRD) for the activation of GUT-2 using a steady heating ramp from 30°C to 100°C over approximately 1 h (**a**) and at a constant temperature of 50°C for 6 h (**b**) in the range of 11.0 to 12.5 degrees. (**c**) Experimental PXRD pattern for the hydration of GUT-2 at RT over 40 h in the range of 11.0 to 12.5 degrees. Panels (**d**)-(**f**) show the corresponding phase quantifications for the activated and hydrated form of GUT-2 that were determined via Rietveld refinements for different temperatures. Here, the data points for the hydrated form are shown as pink circles (○), while the data points for the activated form are represented as purple, inverted triangles (▽).

## S10. Mechanical properties of GUT-2

For infinitesimal deformation strains the crystal energy $\varepsilon$ can be expressed by a quadratic form of the elastic tensor:[12]

$$E = E_0 + \frac{V_0}{2} \sum_{i,j} C_{ij}\varepsilon_i\varepsilon_j + O(\varepsilon^3)$$

A crystal structure is said to be mechanically stable if – in addition to the structure having no negative phonon modes (dynamical stability) – the energy contribution of the quadratic form is always positive. This is mathematically equivalent to the condition that the matrix $C$ is positive definite, i.e., possessing no negative eigenvalues.

As discussed in greater detail in the Supporting Information of Ref.[13] elastic constants are difficult to obtain using density functional theory (DFT). Here, we employ the CRYSTAL23 code[14], which thanks to the use of numerically efficient atom-centered basis sets allows for the calculation of the fully ion-relaxed elastic tensor elements. Regarding the basis sets we used Zn[8-64111-41G(f)][15,16] for the metallic node and C/N/H/O[6-311G(d,p)][16,17] for the linker atoms. Similar to the calculation of the elastic tensor elements for the hydrated structure of GUT-2 in Ref.[13] we set the convergence criteria TOLDEG and TOLDEX to 0.0002 and 0.0004, respectively, and we used the PBE functional[18] in combination with the D3 van der Waals correction.[19] A strain step of 0.05 Å was applied. The computed elastic tensors are provided in Table S15.

**Table S15.** Calculated elastic tensors for hydrated and activated form of GUT-2 using CRYSTAL23.

| Hydrated GUT-2 (taken from Ref.[13]) | Activated GUT-2 |
|---|---|
| $C_{\mathrm{DFT}} = \begin{pmatrix} 19.8 & 12.4 & 8.7 & 0 & 0 & 0 \\ 12.4 & 22.8 & 14.3 & 0 & 0 & 0 \\ 8.7 & 14.3 & 21.4 & 0 & 0 & 0 \\ 0 & 0 & 0 & 4.4 & 0 & 0 \\ 0 & 0 & 0 & 0 & 5.0 & 0 \\ 0 & 0 & 0 & 0 & 0 & 5.2 \end{pmatrix}$ GPa | $C_{\mathrm{DFT}} = \begin{pmatrix} 8.5 & 6.7 & 9.6 & 0 & 0 & 0 \\ 6.7 & 12.3 & 8.7 & 0 & 0 & 0 \\ 9.6 & 8.7 & 19.0 & 0 & 0 & 0 \\ 0 & 0 & 0 & 4.8 & 0 & 0 \\ 0 & 0 & 0 & 0 & 3.5 & 0 \\ 0 & 0 & 0 & 0 & 0 & 4.3 \end{pmatrix}$ GPa |

Although there are three trivial eigenvalues of the matrix $C$ in case of orthorhombic crystal symmetry that need to be positive, namely $C_{44}$, $C_{55}$ and $C_{66}$, it is not possible to give a closed form expression for the Born stability criterion[12] for the remaining three eigenvalues. Hence, we have to calculate them brute-force and check their positivity. All six eigenvalues of the elastic tenors for both structures are listed in Table S16. As can be seen from this table all eigenvalues are strictly positive, hence the Born stability criterion for mechanical stability is fulfilled for both structures.

**Table S16.** Eigenvalues of the elastic tensors for hydrated and activated form of GUT-2 using CRYSTAL23.

| Eigenvalue [GPa] | Hydrated GUT-2 | Activated GUT-2 |
|---|---|---|
| $\lambda_1$ | 4.4 | 2.5 |
| $\lambda_2$ | 4.9 | 3.5 |
| $\lambda_3$ | 5.2 | 4.3 |
| $\lambda_4$ | 6.8 | 4.8 |
| $\lambda_5$ | 12.0 | 6.4 |
| $\lambda_6$ | 45.3 | 31.0 |